\documentstyle[emulateapj]{article}

\newcommand\about     {\hbox{$\sim$}}

\makeatother

\def\non    {\nonumber \\}

\def\case#1/#2{\hbox{$\frac{#1}{#2}$}}
\def\about  {\hbox{$\sim$}}

\def\g                {\hbox{$g^*$}}
\def\r                {\hbox{$r^*$}}

\def\gr               {\hbox{$g^*-r^*$}}

\def\iz               {\hbox{$i^*-z^*$}}

\def\a                {\hbox{$a^*$}}

\def\comment#1        {\tt #1}

\begin{document}


\title{ Comparison of Asteroids Observed in the Sloan Digital Sky Survey$^1$
                      with a Catalog of Known Asteroids                      }

\author{
Mario Juri\'{c}\altaffilmark{\ref{Princeton},\ref{Zagreb}},
\v{Z}eljko Ivezi\'{c}\altaffilmark{\ref{Princeton}},
Robert H. Lupton\altaffilmark{\ref{Princeton}},
Tom Quinn\altaffilmark{\ref{Washington}},
Serge Tabachnik\altaffilmark{\ref{Princeton}},
Xiaohui Fan\altaffilmark{\ref{Princeton}},
James E. Gunn\altaffilmark{\ref{Princeton}},
Gregory S. Hennessy\altaffilmark{\ref{USNO1}},
Gillian R. Knapp\altaffilmark{\ref{Princeton}},
Jeffrey A. Munn\altaffilmark{\ref{USNO2}},
Jeffrey R. Pier\altaffilmark{\ref{USNO2}},
Constance M. Rockosi\altaffilmark{\ref{Chicago}},
Donald P. Schneider\altaffilmark{\ref{PennState}},
Jonathan Brinkmann\altaffilmark{\ref{APO}},
Istv\'an Csabai\altaffilmark{\ref{JHU},\ref{Eotvos}},
Masataka Fukugita\altaffilmark{\ref{CosmicRay},\ref{IAS}}
}

\altaffiltext{1}{Based on observations obtained with the
Sloan Digital Sky Survey.}
\newcounter{address}
\setcounter{address}{2}
\altaffiltext{\theaddress}{Princeton University Observatory, Princeton, NJ 08544
\label{Princeton}}
\addtocounter{address}{1}
\altaffiltext{\theaddress}{University of Zagreb, Dept. of Physics, Bijeni\v{c}ka 
cesta 32, 10000 Zagreb, Croatia
\label{Zagreb}}
\addtocounter{address}{1}
\altaffiltext{\theaddress}{University of Washington, Dept. of Astronomy,
Box 351580, Seattle, WA 98195
\label{Washington}}
\addtocounter{address}{1}
\altaffiltext{\theaddress}{U.S. Naval Observatory,
Washington, DC  20392-5420
\label{USNO1}}
\addtocounter{address}{1}
\altaffiltext{\theaddress}{U.S. Naval Observatory,
Flagstaff Station, P.O. Box 1149, Flagstaff, AZ 86002
\label{USNO2}}
\addtocounter{address}{1}
\altaffiltext{\theaddress}{University of Chicago, Astronomy \& Astrophysics
Center, 5640 S. Ellis Ave., Chicago, IL 60637
\label{Chicago}}
\addtocounter{address}{1}
\altaffiltext{\theaddress}{Dept. of Astronomy and Astrophysics,
The Pennsylvania State University,
University Park, PA 16802
\label{PennState}}
\addtocounter{address}{1}
\altaffiltext{\theaddress}{Apache Point Observatory,
2001 Apache Point Road, P.O. Box 59, Sunspot, NM 88349-0059
\label{APO}}
\addtocounter{address}{1}
\altaffiltext{\theaddress}{Department of Physics and Astronomy, 
The Johns Hopkins University, 3701 San Martin Drive, Baltimore, MD 21218
\label{JHU}}
\addtocounter{address}{1}
\altaffiltext{\theaddress}{Department of Physics of Complex Systems,
E\"otv\"os University, P\'azm\'any P\'eter s\'et\'any 1/A, Budapest, H-1117, Hun
gary
\label{Eotvos}}
\addtocounter{address}{1}
\altaffiltext{\theaddress}{Institute for Cosmic Ray Research, University of
Tokyo, Midori, Tanashi, Tokyo 188-8502, Japan
\label{CosmicRay}}
\addtocounter{address}{1}
\altaffiltext{\theaddress}{Institute for Advanced Study, Olden Lane,
Princeton, NJ 08540
\label{IAS}}

\begin{abstract}
We positionally correlate asteroids from existing catalogs with a sample of
$\about$18,000 asteroids detected by the Sloan Digital Sky Survey (SDSS, 
Ivezi\'{c} {\em et al.} 2001). We find 2641 unique matches, which represent 
the largest sample of asteroids with both accurate multi-color photometry and 
known orbital parameters. The matched objects are predominantly bright, and 
demonstrate that the SDSS photometric pipeline recovers \about90\% of the 
known asteroids in the observed region. For the recovered asteroids we find 
a large offset (\about 0.4 mag) between Johnson V magnitudes derived from 
SDSS photometry and the predicted catalog-based visual magnitudes. This offset 
varies with the asteroid color from 0.34 mag for blue asteroids to 0.44 mag for 
red asteroids, and is probably caused by the use of unfiltered CCD observations 
in the majority of recent asteroid surveys. This systematic photometric error 
leads to an overestimate of the number of asteroids brighter than a given 
absolute magnitude limit by a factor of \about 1.7. The distribution of the 
matched asteroids in orbital parameter space indicates strong color segregation. 
We confirm that some families are dominated by a single asteroid type (e.g. the 
Koronis family by red asteroids and the Themis family by blue asteroids), while 
others appear to be a mixture of blue and red objects (e.g. the Nysa/Polana family). 
Asteroids with the bluest \iz\ colors, which can be associated with the Vesta 
family, show particularly striking localization in orbital parameter space.
\end{abstract}

\section{                     Introduction               }

The Sloan Digital Sky Survey (SDSS) is a digital photometric and spectroscopic
survey which will cover 10,000 deg$^2$ of the Celestial Sphere in the North Galactic
cap and produce a smaller ($\sim$ 225 deg$^2$) but much deeper survey in the
Southern Galactic hemisphere (York {\em et al.} 2000 and references therein).
The survey sky coverage will result in photometric measurements for about
50 million stars and a similar number of galaxies. The flux densities of detected
objects are measured almost simultaneously in five bands ($u$, $g$, $r$, $i$,
and $z$; \cite{F96}) with effective wavelengths of 3551 \AA, 4686 \AA, 6166 \AA,
7480 \AA, and 8932 \AA, 95\% complete\footnote{These values are determined by
comparing multiple scans of the same area obtained during the commissioning year.
Typical seeing in these observations was 1.5$\pm$0.1 arcsec.} for point sources to
limiting magnitudes of 22.0, 22.2, 22.2, 21.3, and 20.5 in the North Galactic
cap\footnote{We refer to the measured magnitudes in this paper as $u^*, g^*, r^*,
i^*,$ and $z^*$ because the absolute calibration of the SDSS photometric system
(dependent on a network of standard stars) is still uncertain at the $\sim 0.03^m$
level. The SDSS filters themselves are referred to as $u, g, r, i,$ and $z$.
All magnitudes are given on the AB$_\nu$ system (Oke \& Gunn 1983, for additional
discussion regarding the SDSS photometric system see \cite{F96}, and Stoughton
{\em et al.} 2002).}. Astrometric positions are accurate to about 0.1 arcsec
per coordinate (rms) for sources brighter than 20.5$^m$ (Pier {\em et al.} 2001),
and the morphological information from the images allows robust star-galaxy
separation to $\sim$ 21.5$^m$ (Lupton {\em et al.} 2002).

SDSS, although primarily designed for observations of extragalactic objects, will
significantly contribute to studies of the solar system objects, because asteroids
in the imaging survey must be explicitly detected to avoid contamination of the 
samples of extragalactic objects selected for spectroscopy. Ivezi\'{c} {\em et al.}
(2001, hereafter I01) analyzed SDSS commissioning data and showed that SDSS will
increase the number of asteroids with accurate five-color photometry by more than
two  orders of magnitude (to about 100,000), and to a limit about seven magnitudes
fainter than previous multi-color surveys (e.g. The Eight Color Asteroid Survey
by Zellner, Tholen \& Tedesco 1985). The main results derived from these early SDSS
observations are

\begin{enumerate}
\item
A measurement of the main-belt asteroid size distribution to a significantly smaller
size limit (\about1 km) than possible before. The size distribution resembles
a broken power-law, independent of the heliocentric distance: $D^{-2.3}$ for 0.4 km
$\la D \la$ 5 km, and $D^{-4}$ for 5 km $\la D \la$ 40 km.
\item
A smaller number of asteroids compared to previous work. In particular,
the number of asteroids with diameters larger than 1 km is about $7\times10^5$,
or up to three times less then suggested by earlier studies. 
\item
The distribution of main-belt asteroids in 4-dimensional SDSS color
space is strongly bimodal, and the two groups can be associated with S (rocky)
and C (carbonaceous) type asteroids. A strong bimodality is also seen in
the heliocentric distribution of asteroids: the inner belt is dominated by
S type asteroids centered at $R$ \about 2.8 AU, while C type asteroids,
centered at $R$ \about 3.2 AU, dominate the outer belt. 
\end{enumerate}

I01 demonstrated that the SDSS photometric pipeline ({\em photo},
Lupton {\em et al.} 2002) is a robust and highly efficient automated
tool for finding moving objects. I01 present a detailed discussion of
the completeness and reliability of the SDSS asteroid catalog based on the known
SDSS photometric and astrometric precision, but this discussion is incomplete
because it does not attempt a direct comparison with catalog of known asteroids 
on an object-by-object basis. The importance of such an analysis is evident from 
the realization that there are \about 90,000 {\em cataloged} asteroids with $H < 15.5$, 
while I01 expect to find only \about 58,000 in the same absolute magnitude\footnote{The 
absolute magnitude is the magnitude that an asteroid would have at a distance of 
1 AU from the Sun and from the Earth, viewed at zero phase angle. For more 
details see I01 and references therein.} range. The implied low completeness of 
64\% (compare to 98\% estimated by I01) can only be verified by direct matching of 
known and SDSS asteroids\footnote{We thank R. Jedicke for pointing out this discrepancy 
to us.}.

Another motivation for cross-correlating SDSS asteroids and known asteroids
is a large potential increase in the number of asteroids with both
accurate multi-color photometry and known orbital parameters (the SDSS
data themselves are insufficient for accurate orbit determination, but will provide
serendipitous photometric measurements for a substantial fraction of
known asteroids). Additionally, the SDSS color information may be
utilized to study the chemical segregation in the full orbital parameter
space, rather than only as a function of heliocentric distance as
done by I01.

This paper presents the first results on cross-correlating SDSS asteroids
and cataloged asteroids. Section 2 describes the analyzed data, a software
pipeline used for generating the positions of known asteroids at the time of
SDSS observations, and their positional matching to objects automatically
detected by the SDSS photometric pipeline. Section 3 discusses the statistics
of matched objects, and compares the SDSS photometric measurements
with the apparent magnitudes predicted from cataloged absolute magnitudes.
In Section 4 we discuss the chemical segregation of asteroids in orbital element 
space, and in Section 5 we summarize the main results.

\section{   The Matching of SDSS Moving Objects and Known Asteroids    }

\subsection{                    SDSS  data                       }
\label{sdssdata}
We use asteroid data (c.f. I01) from the SDSS Early Data Release, described in 
detail by Stoughton {\em et al.} 2002 (hereafter SDSSEDR). These data include 
equatorial observing runs 94, 125, 752 and 756 (see I01 and SDSSEDR); the boundaries 
are given by $-1.27^\circ \la {\rm Dec} \la 1.27^\circ$, and $RA$ = 351$^\circ$ --
56$^\circ$ (runs 94 and 125), or $RA$ = 145$^\circ$ -- 250$^\circ$ (runs 752
and 756). For a footprint in ecliptic coordinates, see Figure 1 in I01. This region 
has an area of 432 deg$^2$ and includes 12,668 moving objects selected as in I01 \
(the velocity $v > 0.03$ deg/day and $14.0 < r^* < 21.5$). We use this sample for
all quantitative estimates of the catalog completeness, and for photometric 
comparison. In order to increase the number of objects when studying the color 
segregation in the asteroid belt (Section \ref{colseg}), we also add additional
objects from seven currently unreleased equatorial observing runs\footnote{Prompt 
release of asteroid data (i.e. without any waiting period) is currently under 
consideration by SDSS.} that roughly double the matched sample size.

The SDSS photometric data include a list of objects flagged as moving by the photometric
pipeline. For each object, the position and time of observation,
and the magnitudes and associated errors in five SDSS photometric bands ($u, g, r,
i, z$) are recorded. The SDSS imaging data is obtained in the time-delay-and-integrate
(TDI, or ``drift-scanning") mode\footnote{See http://www.astro.princeton.edu/PBOOK/welcome.htm},
and thus each observed position corresponds to a different observing time (as opposed 
to the starring mode where all objects from a given image are observed at the
same time). In a general case,
the correlation between a position and time is easiest to compute in the great
circle coordinate system. Since all the scans discussed here were obtained along
the Celestial Equator, this dependence becomes particularly simple, and is given by
\begin{equation}
\label{Tdrift}
       T(RA) = T_0 + {RA - RA_0 \over 360^\circ} {\rm days},
\end{equation}
where $T$ is the time corresponding to position $RA$ (e.g. Julian Day), and the 
zeropoints $T_0$ and $RA_0$ are constants for a given run (all reported positions 
correspond to the $r$ band). For an arbitrary run, this expression is easily 
generalized by using an appropriate coordinate system. 

The cataloged asteroid magnitudes are reported in the Johnson $V$ band. Preliminary 
transformations from the SDSS photometric system to the Johnson bands
are given by Fukugita {\em et al.} (1996) and Krisciunas, Margon \& Szkody (1998).
We use a recent updated version of these transformations (M. Fukugita, priv. comm.;
the new transformations produce V magnitudes that agree to better than 0.1 mag with 
the version from Fukugita {\em et al.} 1996) to synthesize the Johnson B and V band 
magnitudes\footnote{Although SDSS has already produced more
multi-color photometric measurements of asteroids than is available in the Johnson
system, we use the Johnson system to ease comparison with earlier studies.}
\begin{eqnarray}
             V_o = \r + 0.44 \, (\g-\r) - 0.02   \\
         (B-V)_o =  1.04 \, (\g-\r) + 0.19      \non
\end{eqnarray}

For typical values of \gr\ for asteroids (0.4--0.8), $V_o-\r$ is in the range 0.16 to 0.33.
The overall accuracy of these transformations and SDSS photometry is better than
\about 0.05 mag, as determined by direct comparison with non-SDSS observations
obtained in the Johnson system (M. Fukugita, E. Grebel, J. Holtzman, unpublished).

\subsection{         The catalog of known asteroids                }

For positional matching of the moving objects observed by SDSS with
cataloged asteroids we select the Asteroid Orbital Elements Database
(ASTORB). ASTORB is a catalog of high-precision osculating orbital elements and
other information on all numbered and a large number of unnumbered asteroids
(but does not contain information on known numbered comets). The ASTORB catalog is
distributed by the Lowell
Observatory in the form of an ASCII file, containing single line records for
each asteroid (Bowell 2001).  The catalog is
updated daily for addition of newly discovered objects, deletion of duplicate objects, 
and improvement of parameters of known objects. In this work we use
the six osculating orbital elements, absolute magnitude and the phase-correction
slope parameter which are necessary to predict the asteroid position and
apparent magnitude at the time of SDSS scans. We also utilize the arc (in days)
spanned by the observations used to compute the orbit to estimate the accuracy
of orbital elements.

Orbital elements given in ASTORB are heliocentric and have been computed by
a variable-timestep differential orbit correction algorithm, based on astrometric
observations obtained from the Minor Planet Center. The perturbations from all
the major planets, the Moon and the three largest asteroids (Ceres, Pallas and
Vesta) have been taken into account. Absolute magnitude corresponding to the
Johnson $V$ band ($H$, see I01) is listed as numbers rounded to two, one, or no
decimal places, with the number of decimal
places reflecting  assumed reliability. However, for unnumbered asteroids $H$
is given to two decimal places regardless of its real accuracy. The slope parameter
$G$ (see section \ref{apmags} below) is given for asteroids for which it is known, 
and for others, which are the overwhelming majority of the sample, assumed to be 
0.15 (dimensionless).

We note that besides ASTORB there are several other asteroid orbital
elements databases (e.g. Minor Planet Center Orbit Database\footnote{Available
from ftp://cfa-ftp.harvard.edu/pub/MPCORB}, Asteroid Dynamics Site\footnote{Available
from http://hamilton.dm.unipi.it/cgi-bin/astdys/astibo}).
However, because of its widespread use and significant additional information
supplied for each asteroid, we have chosen to use ASTORB as a referent catalog
for this study. The ASTORB version used here is from September 18, 2001, and
contains 141,110 objects (29,074 of which are numbered asteroids).

\subsection{          Asteroid identification pipeline          }

The asteroid identification pipeline consists of three parts:

\begin{enumerate}
\item Propagation of the asteroid osculating orbital elements from ASTORB 
      to the epoch of the SDSS observation
\item Computation of the asteroid positions and apparent magnitudes at the time
      of SDSS scan
\item Positional matching of SDSS moving objects with the known asteroids
\end{enumerate}

\subsubsection{              Propagation of orbits                  }

The ASTORB catalog contains osculating orbital elements computed for epoch near
the current epoch, where ``current'' corresponds to the publishing date of
the ASTORB catalog file. The orbital elements are propagated to the epoch of
observation using the PROELE routine of OrbFit v1.8 {\em propag} library (\cite{Milani99}).
We use the default dynamical model supplied with the OrbFit package. It includes
gravitational perturbations of the Sun, Moon and the major planets and
relativistic corrections, when necessary. The positions and masses for major
planets are derived from JPL ephemeris DE405 (Standish 1998). Although the OrbFit package
offers multiple choices for the numerical integration schemes, we use its automated
selection procedures, which has been proven to be sufficiently accurate in practice 
(\cite{Juric00}).
The end result of the orbit propagation is a catalog of osculating orbital elements
at the time of SDSS observation, which are then used for quick two-body (Sun and 
asteroid) computations of the ephemeris.

\subsubsection{ Computation of asteroid positions and apparent magnitudes    }
\label{apmags}

The calculations of positions are performed using a modified PREOBS routine from
the same OrbFit package (the modification consists of enabling a two-body
approximation). When calculating the positions of cataloged asteroids, they must
satisfy the condition given by eq.~\ref{Tdrift}. We solve the problem iteratively,
until the difference between the positions calculated in two successive iterations
is smaller than 0.001 arcsec. In practice the computations converge rapidly and
usually reach the required accuracy with two to three iterations.

The apparent magnitudes are computed by OrbFit's APPMAG subroutine from the cataloged
values of the absolute magnitude $H$ and the phase-correction slope parameter $G$.
The phase-corrected apparent visual ``catalog'' magnitude, $V_c$,  is obtained from
(Bowell 1999)
\begin{equation}
   V_c(\alpha)  = H + 5 \log(R \Delta) - 2.5 \log((1-G)\Phi_1(\alpha) + G \Phi_2(\alpha))
\end{equation}
where $R$ is the heliocentric, and $\Delta$ is the geocentric distance (expressed
in AU), $\alpha$ is the ``solar'' angle (the angle between the Sun and the Earth as
viewed from the asteroid, see I01), and $\Phi_1$ and $\Phi_2$ are the ``phase-correction''
functions. The latter are obtained from standard approximations
\begin{eqnarray}
      \Phi_{i} = {\rm exp}\left[{-A_i \left(\tan({\alpha \over 2})\right)^{B_i}}\right];
            \,\, i = 1, 2 \\
        A_1 = 3.33\ \,\ A_2 = 1.87  \non
        B_1 = 0.63\ \,\ B_2 = 1.22 \non
\end{eqnarray}

Note that these approximations are formally different from the one used by I01 (eq. 11).
However, they are similar numerically; the difference is less than 0.05 mag
for $\alpha < 2^\circ$ and $\alpha \about 6^\circ$, with the maximum
value of \about0.15 mag at $\alpha \about 3^\circ$.

\subsection{              Matching algorithm          }

The matching algorithm compares the list of predicted positions for the
known asteroids with the list of SDSS objects flagged as moving by the photometric
pipeline. After finding the nearest-neighbor pairs from the two lists, it
requires that the distance between the two positions is less than 30
arcsec. The final sample is very insensitive to the precise value of
this cut because for the majority (86\%) of matched objects the distance
between the two positions is less than 3 arcsec (centered on zero in both
coordinates). The probability that a randomly chosen position would fall within 
3 arcsec from a moving SDSS object is of the order 10$^{-4}$ in the regions with 
the highest asteroid number density (64 deg$^{-2}$, I01), implying that only about 
one of the matches within 3 arcsec is a random association. The associated objects 
are taken to be positive positional identifications of an SDSS moving object with 
a cataloged asteroid.

\section{                The Matching  Results                          }

The SDSS photometric pipeline identified 12,668 moving objects selected as
described by I01. There are 2801 cataloged asteroids whose computed positions
are within the boundaries of the SDSS scans considered here, and 1633 have
within 30 arcsec an object flagged by SDSS photometric pipeline as moving. Of 
these, only 1325 are unique objects because some asteroids were observed by SDSS 
more than once (I01). Detailed statistics for each run are listed in Table 1.

This comparison, taken at face value, implies that the SDSS sample is 58\%
complete. This is a much lower completeness than claimed by I01 (98\%). 
We have visually inspected SDSS images around the positions of all 1168 ASTORB 
objects not found in the SDSS catalog (moving objects are easily recognized on 
1x1 arcmin g-r-i color composites thanks to their peculiar appearances, see I01). 
There are 219 objects which are clearly moving, but were not recognized as such
by the photometric pipeline. The most common reason for missing them
are complex environments (bright stars with diffraction spikes, compact galaxy
clusters, meteor and airplane trails, etc.) which are hard to ``deblend''
into individual sources. The revised, true completeness of the SDSS
sample is thus 88\%, somewhat lower than 98\% determined by I01. We note that 
I01 determined completeness using only data from run 756. For this run the 
completeness determined here is 91\%. Furthemore, for all runs except for run 
94 the completeness is higher than 90\%. The overall completeness is below 90\%
only for run 94 (81\%), which is one of the earliest SDSS commissioning runs, 
obtained while the telescope still did not have proper optics.

For the remaining 949 (34\% of the total) ASTORB objects there is no visible
SDSS source within 30 arcsec from the predicted position down to the sensitivity
limit of \r \about 22.5. The most plausible explanation for these ``missing''
sources is inaccurate orbital elements (that is, their true position
during SDSS observations is further than 30 arcsec from the predicted
position). Indeed, the inspection of their entries in the ASTORB catalog
indicates that the majority are either faint ($H \ga 17$), have a small 
observational arc used to determine the orbital elements, or both.

\subsection{  The selection of asteroids with reliable orbital elements  }

SDSS observations demonstrate that a large fraction of known asteroids listed
in the ASTORB catalog do not have sufficiently accurate orbital elements for
reliable identification. The observational arc used to determine the orbital
elements can be utilized as a rough estimate of their quality -- the longer
the arc spanned by the observations used to compute the elements, the more
accurate are the positions. The ASTORB catalog also provides a more quantitative
estimate for the accuracy of the orbital elements -- the \emph{Current ephemeris
uncertainty ($CEU$)} parameter (1-$\sigma$ absolute positional uncertainty),
for an epoch near the catalog publishing date. However, the $CEU$ cannot be 
propagated to the time of observation without the knowledge of covariance matrices 
for the orbital elements solution (cf. \cite{MB93}), and we use arc to make a 
quality cut.

\begin{figure}
\plotone{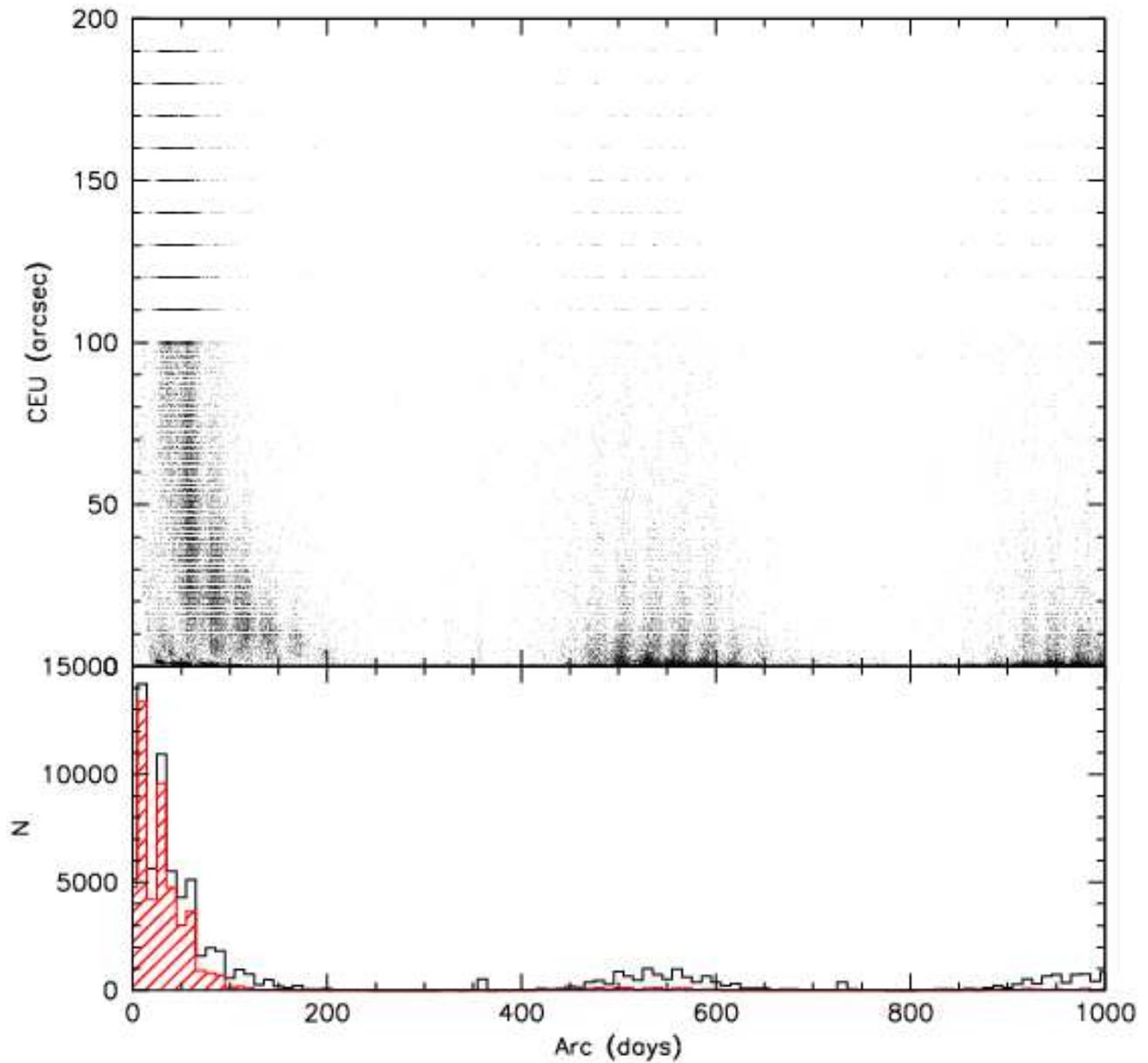}
\caption{
The relation between the observational arc and the current ephemeris
uncertainty ($CEU$) for the 141,110 objects from the ASTORB catalog,
marked by dots. The discretization is caused by truncation. The bottom panel 
shows the arc histograms for
all objects, and separately for objects with $CEU > $ 30 arcsec, shown
by the shaded area. As evident, most of the latter are concentrated in
the first peak and are eliminated by the condition arc $>$ 300 days.
\label{fig_arc}
}
\end{figure}

The relationship between the $CEU$ and the arc is shown in Figure
\ref{fig_arc}. The top panel shows each object from the ASTORB catalog
as a dot, and the bottom panel shows the arc histograms for all objects by
a full line, and separately for objects with $CEU > $ 30 arcsec as a shaded
region. As evident, the requirement that arc $>$ 300 days successfully eliminates most
of the asteroids with large CEUs, and selects 1634 objects (out of 2801).
Note that the resulting sample is insensitive to the precise value of this cutoff
because of the non-continuous distribution of data.
A further constraint is based on the predicted apparent magnitude. We only
select asteroids that are reliably detectable by SDSS; they satisfy
$14 \le V_c \la 21.5$  (I01). This cut has only a minor importance and
selects 1612 objects (relaxing the limit to 22.5 adds 3 objects).

\begin{figure}
\plotone{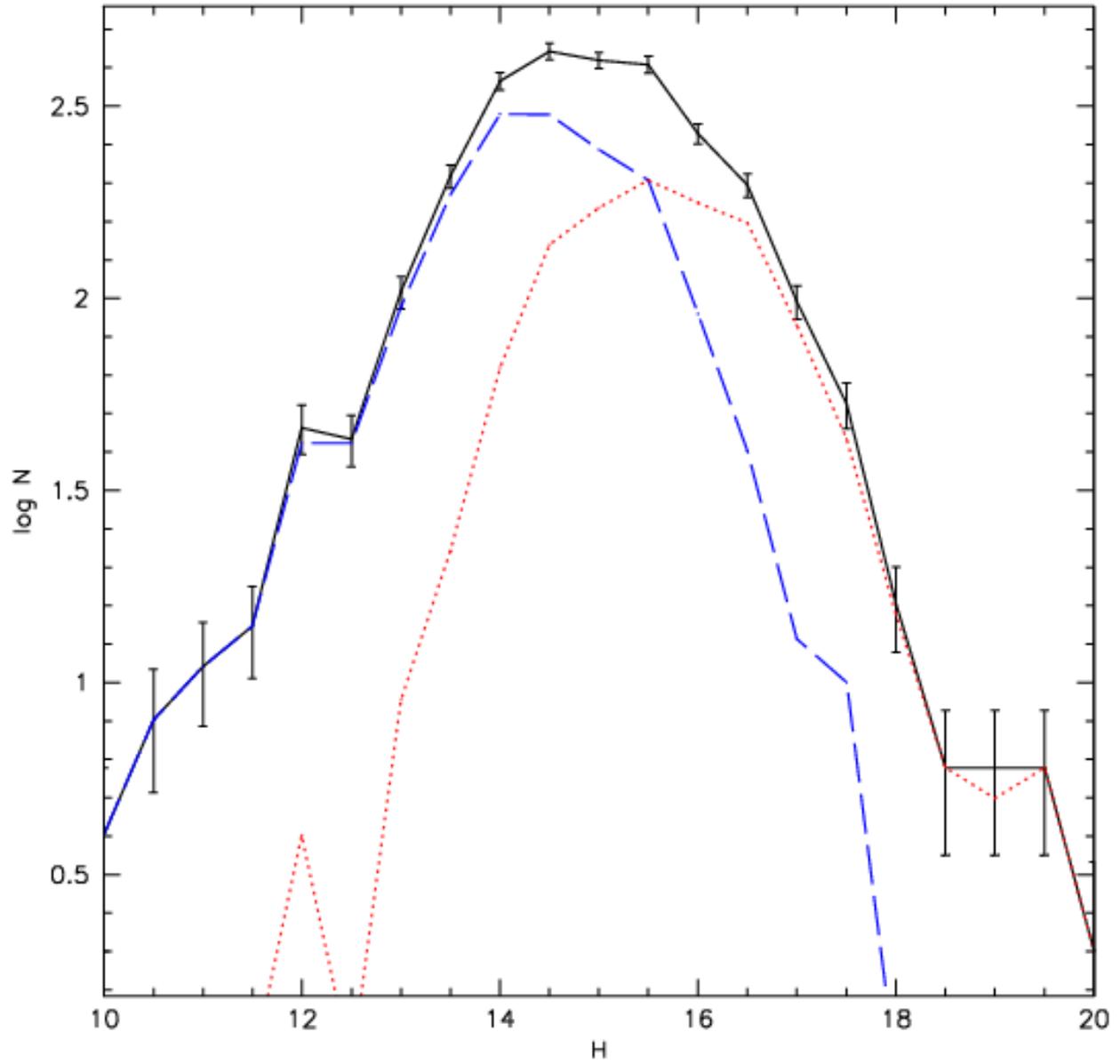}
\caption{The solid line shows the absolute magnitude distribution for
the 2801 known asteroids from the ASTORB database which are expected to be 
observed in SDSS scans discussed here. The distribution for the 1612 asteroids 
with most reliable orbital elements is shown by the dashed line, and the dotted 
line show the distribution for the remaining sources with less reliable orbital
elements.
\label{fig_hlog}
}
\end{figure}

The absolute magnitudes for selected (hereafter ``reliable'') and removed
asteroids (hereafter ``unreliable'') are compared in Figure \ref{fig_hlog}.
The solid line shows the absolute magnitude distribution for all 2801 asteroids,
and the dashed and dotted lines show the distributions for the 1612 (58\%)
reliable and 1189 (42\%) unreliable asteroids, respectively. It is evident that
the asteroids with unreliable orbital parameters are mostly faint, and dominate
the sample for $H > 15.5$. The subsample of reliable asteroids appears complete
for $H < 14$.

The removal of the sources with unreliable orbital elements is efficient,
but not perfect. The matching statistics for the subsample of ASTORB asteroids
with reliable orbits is listed in Table 2. Out of 1612 objects, 1335 are matched
within 30 arcsec to objects automatically recognized by the photometric
pipeline as moving. Of the remaining 277 objects, 173 are visually identified
as moving, and the remaining 104 objects do not have an SDSS moving source within
30 arcsec. For each of these 104 objects, the predicted position has been
independently verified using the on-line AstDys orbital
calculator\footnote{Available at http://hamilton.dm.unipi.it/cgi-bin/astdys/astibo}.
In summary, the arc $>$ 300 days cut decreases the fraction of ASTORB catalog
asteroids with unreliable orbital elements from 34\% to 6\%.
The SDSS completeness based on the reduced sample is 89\% (=1335/1508), consistent
with the estimate based on the full sample (88\%). Since the removed objects
are predominantly faint, this agreement indicates that the SDSS
completeness is not strongly dependent on  apparent
magnitude.

\subsection{               The Matching Statistics               }

The sample of 1335 matched objects that were detected by the photometric
pipeline, together with the sample of 173 visually confirmed objects from the 
ASTORB catalog, can be used to determine whether the completeness of the SDSS automatically
detected sample depends on magnitude. The top panel in Figure \ref{fig_hlog_det}
compares the absolute magnitude distribution, as listed in the ASTORB catalog,
for the 1335 matched objects (dashed line) to the distribution of all 1612
objects (full line). The bottom panel shows the fraction of matched sources
in each magnitude bin, shown by full line with error bars. The horizontal line
is added to guide the eye and represents the overall completeness of 89\%.
As evident, there is no significant correlation between the
fraction of matched sources and the magnitude.

We have also tested for correlations between the fraction of matched sources
and the phase angle, the object's color and apparent magnitude and did
not find any significant dependences.

\subsubsection{       The apparent magnitude distribution           }

\begin{figure}
\plotone{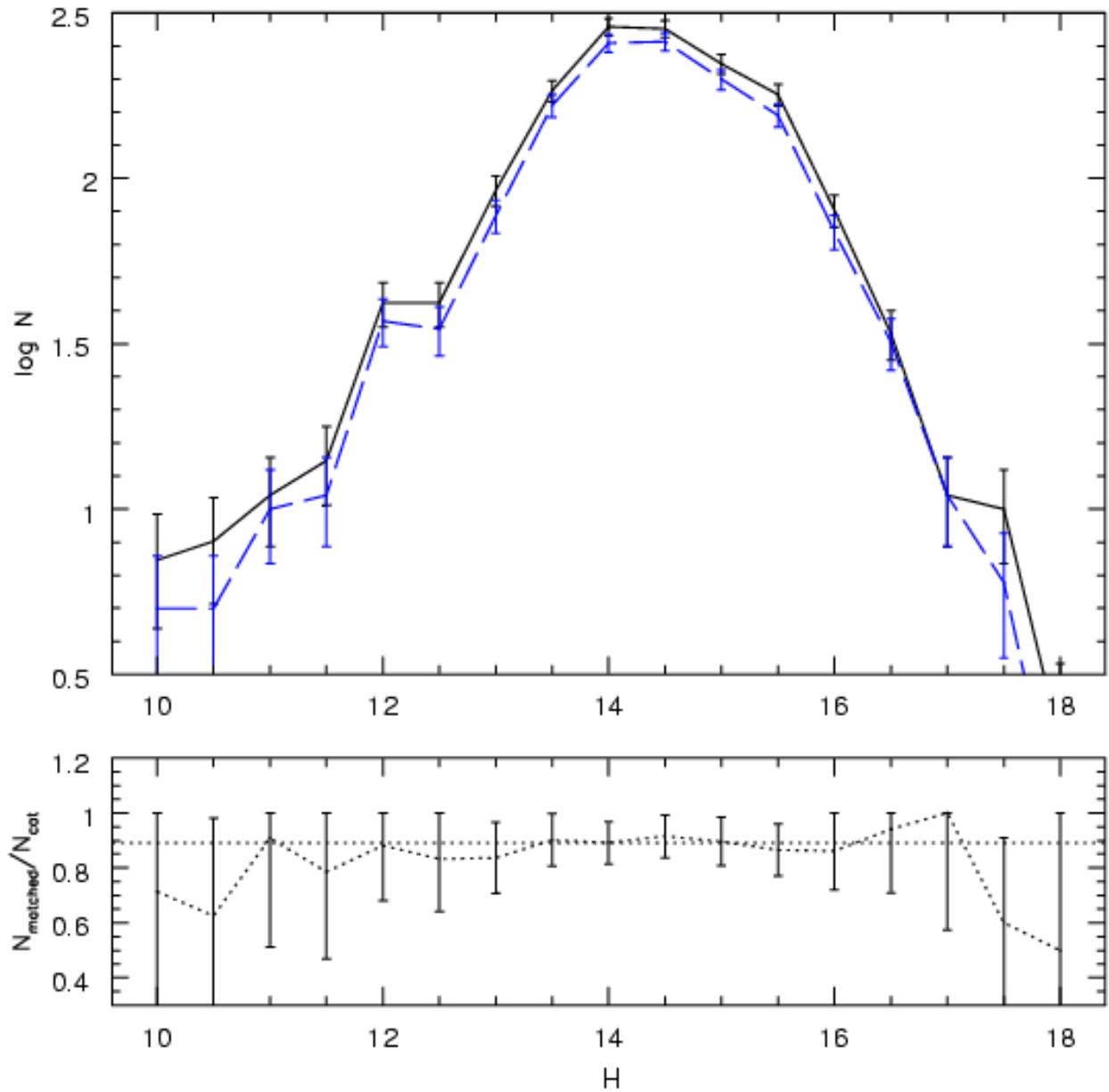}
\caption{The top panel shows the absolute magnitude distribution, as listed
in the ASTORB catalog, for the 1335 objects recognized as moving by the
SDSS photometric pipeline and matched within 30 arcsec to an object from
the ASTORB catalog (dashed line), and the distribution of all 1508
moving objects selected from the ASTORB catalog (full line). The bottom panel
shows the fraction of matched sources in each magnitude bin, shown by full
line with error bars. The horizontal line is added to guide the eye and represents
the overall SDSS completeness of 89\%. Note that there is no
significant correlation between the fraction of matched sources and the
magnitude.
\label{fig_hlog_det}
}
\end{figure}

\begin{figure}
\plotone{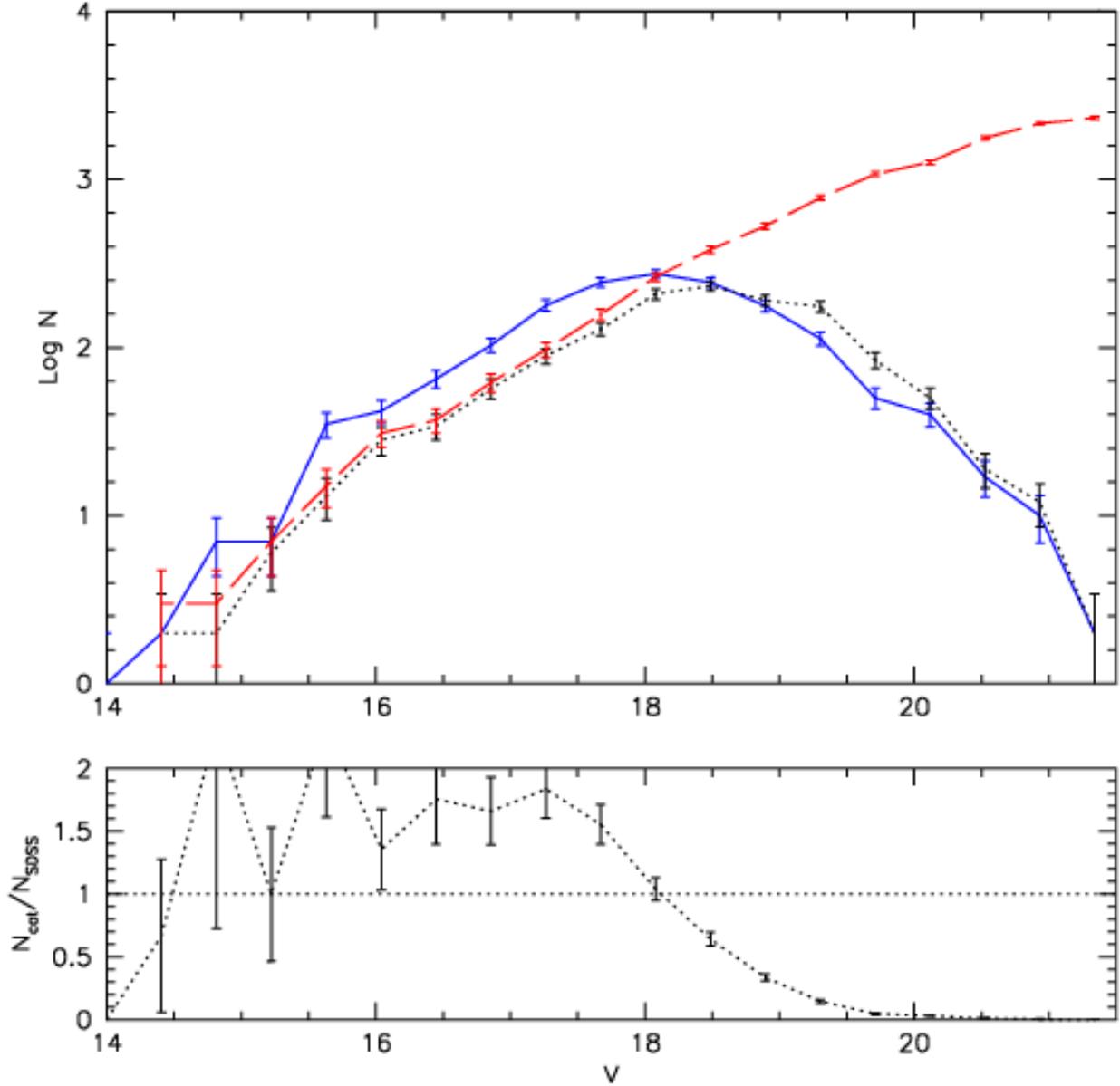}
\caption{The top panel compares the apparent magnitude distribution for asteroids
from the ASTORB catalog that are within the boundaries of SDSS scans discussed here
(solid line), with the apparent magnitude distributions for moving objects detected
by SDSS. The distribution for all SDSS objects is shown by the dashed line and the
distribution for matched objects by the dotted line. The bottom panel shows the
number ratio of ASTORB and SDSS asteroids in each magnitude bin. Although it appears
that the SDSS completeness is as low as 60\% for $V \la 17.5$, the SDSS catalog
includes 89\% of ASTORB asteroids, and the mismatch is due to an offset in the apparent
magnitude scales.
\label{fig_vlog}
}
\end{figure}

The top panel in Figure \ref{fig_vlog} compares the apparent magnitude
distribution for asteroids from the ASTORB catalog that are within the
boundaries of SDSS scans discussed here (full line), with the apparent
magnitude distributions for moving objects detected by SDSS. The distribution
for all SDSS objects is shown by the dashed line and the distribution for
matched objects by the dotted line. We use the calculated $V_c$ magnitudes
for the ASTORB asteroids and synthesized $V_o$ magnitudes for SDSS
asteroids (including the matched sample; this is why the dotted line is above
the solid line for $V \ga 19$). At first sight it appears that
the SDSS sample is significantly incomplete. The bottom panel shows the number ratio
of ASTORB and SDSS asteroids in each magnitude bin, and indicates that the
SDSS completeness is as low as 60\% for $V \la 17.5$. However, as demonstrated
in the previous section (see Figure \ref{fig_hlog_det}), the SDSS catalog
{\em includes} 89\% of ASTORB asteroids, and the mismatch in Figure \ref{fig_vlog}
is simply due to an offset in apparent magnitude scales.

\begin{figure}
\plotone{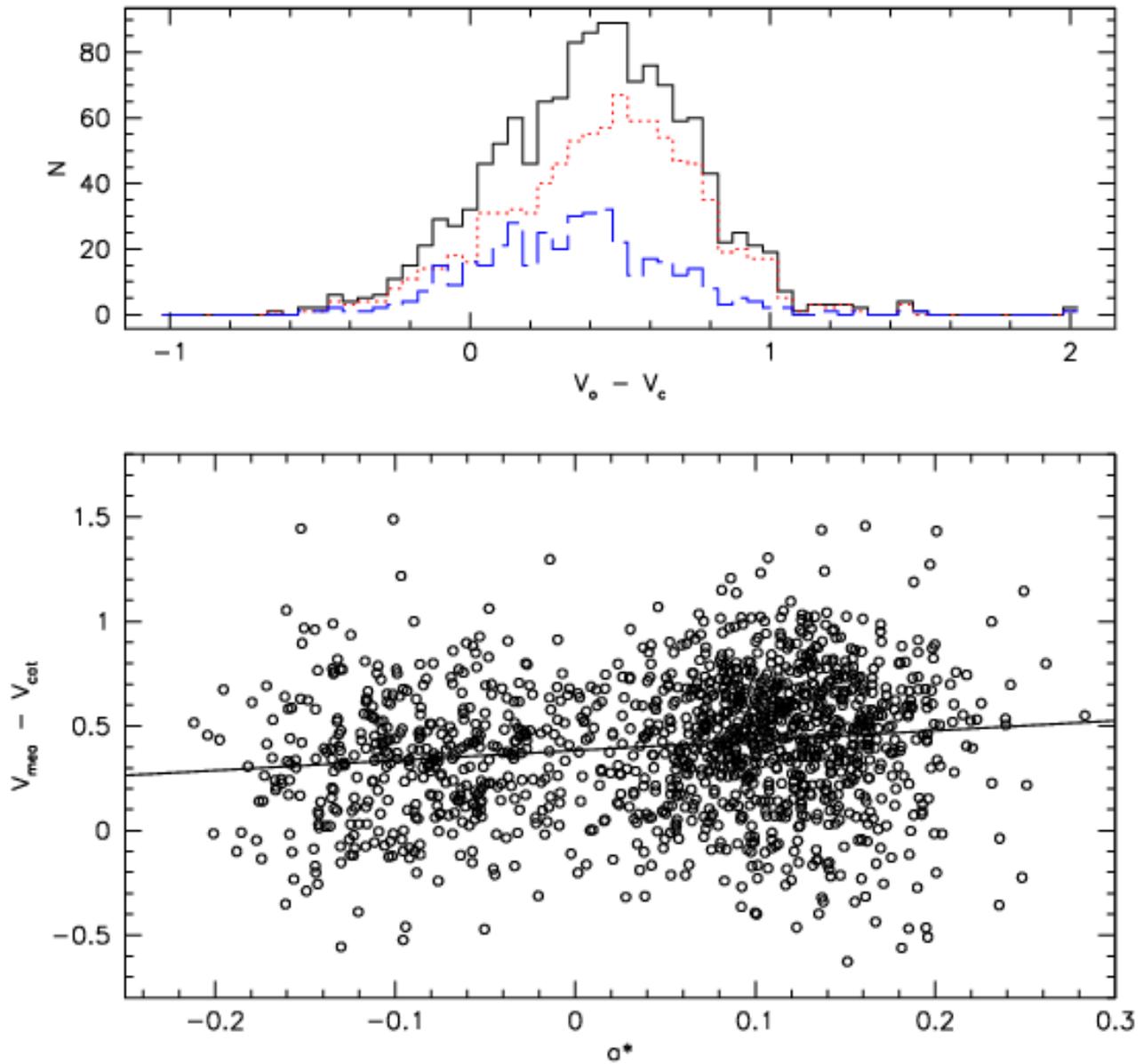}
\caption{The $V_o - V_c$ histogram for all 1335 objects is shown in the top
panel by the full line. The dotted and dashed lines show the analogous histograms
for the 400 blue asteroids and the 935 red asteroids. The bottom panel shows
the $V_o - V_c$ difference as a function of the asteroid color \a, and a best
linear fit.
\label{fig_mdiff_hist}
}
\end{figure}

\begin{figure}
\plotone{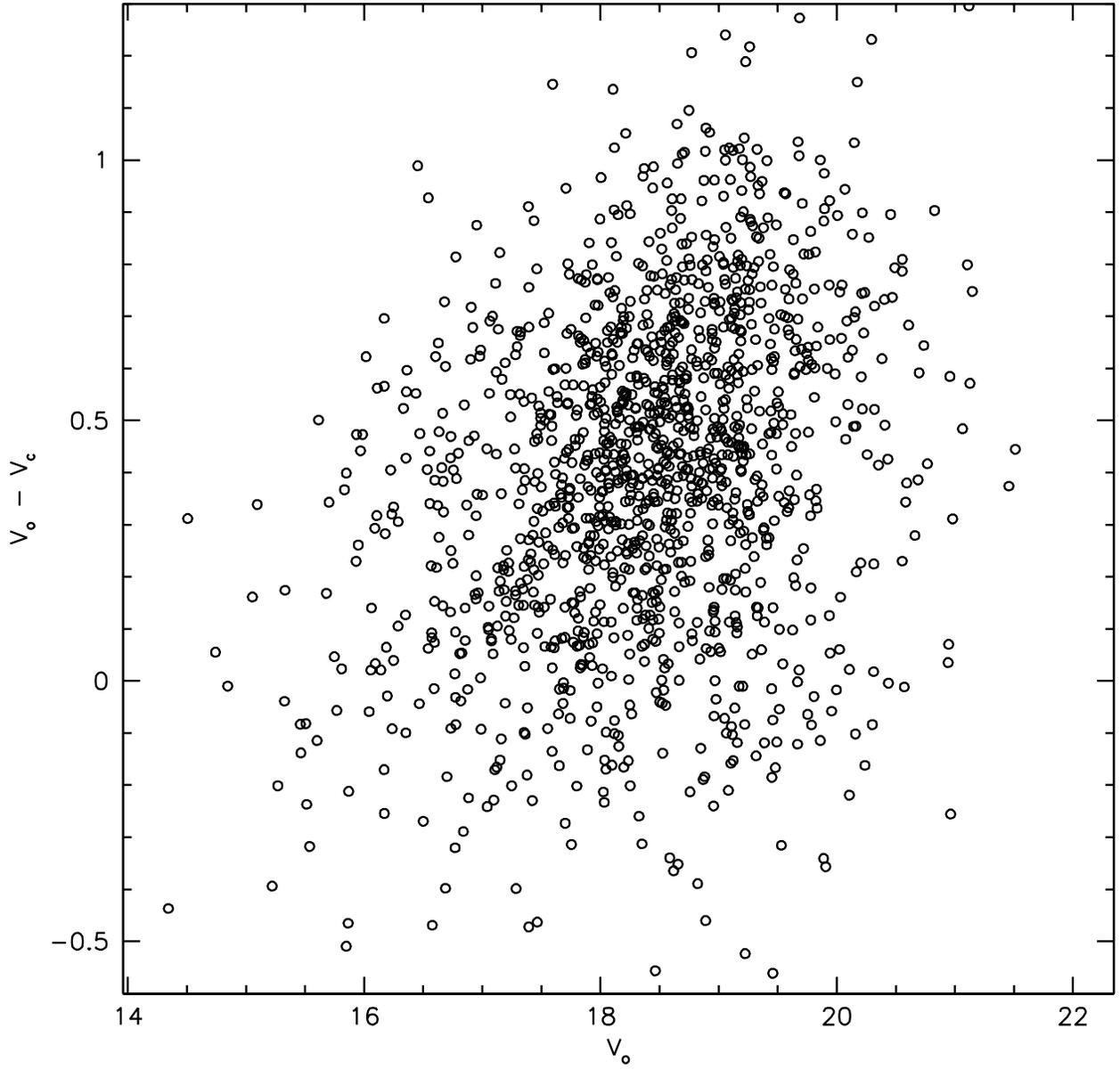}
\caption{The $V_o - V_c$ difference for all 1335 objects as a function of
$V_o$.
\label{fig_mdiff}
}
\end{figure}

The existence of such an offset is further corroborated by the direct
comparison of the calculated ($V_c$, based on the ASTORB catalog) and
synthesized ($V_o$, based on SDSS observations) magnitudes for the 1335
matched objects. The top panel in Figure \ref{fig_mdiff_hist}
shows the histogram of the difference $V_o - V_c$ for all 1335 objects
by the full line. The median offset is 0.41 mag (the SDSS values
are fainter), with the root-mean-scatter of 0.35 mag. The dotted and dashed
lines show the $V_o - V_c$ histograms for the 400 blue asteroids and the 935
red asteroids selected by their \a\ color as described in I01. It is evident
that the apparent magnitude offset depends on the asteroid color; the median
offsets are 0.34 and 0.44 for the blue and red subsample, respectively.
This dependence is further illustrated in the bottom panel, where
the $V_o - V_c$ difference is plotted as a function of the asteroid
color \a. The $V_o - V_c$ histograms shown in Figure \ref{fig_mdiff_hist}
are not symmetric and give a hint of two components: one centered at \about
0.1 is independent of color, and the other one centered at \about 0.4-0.5 \
that appears to be color-dependent. This bimodality is more pronounced for
$V \la 17$, as discernible in Figure \ref{fig_mdiff}.

\begin{figure}
\plotone{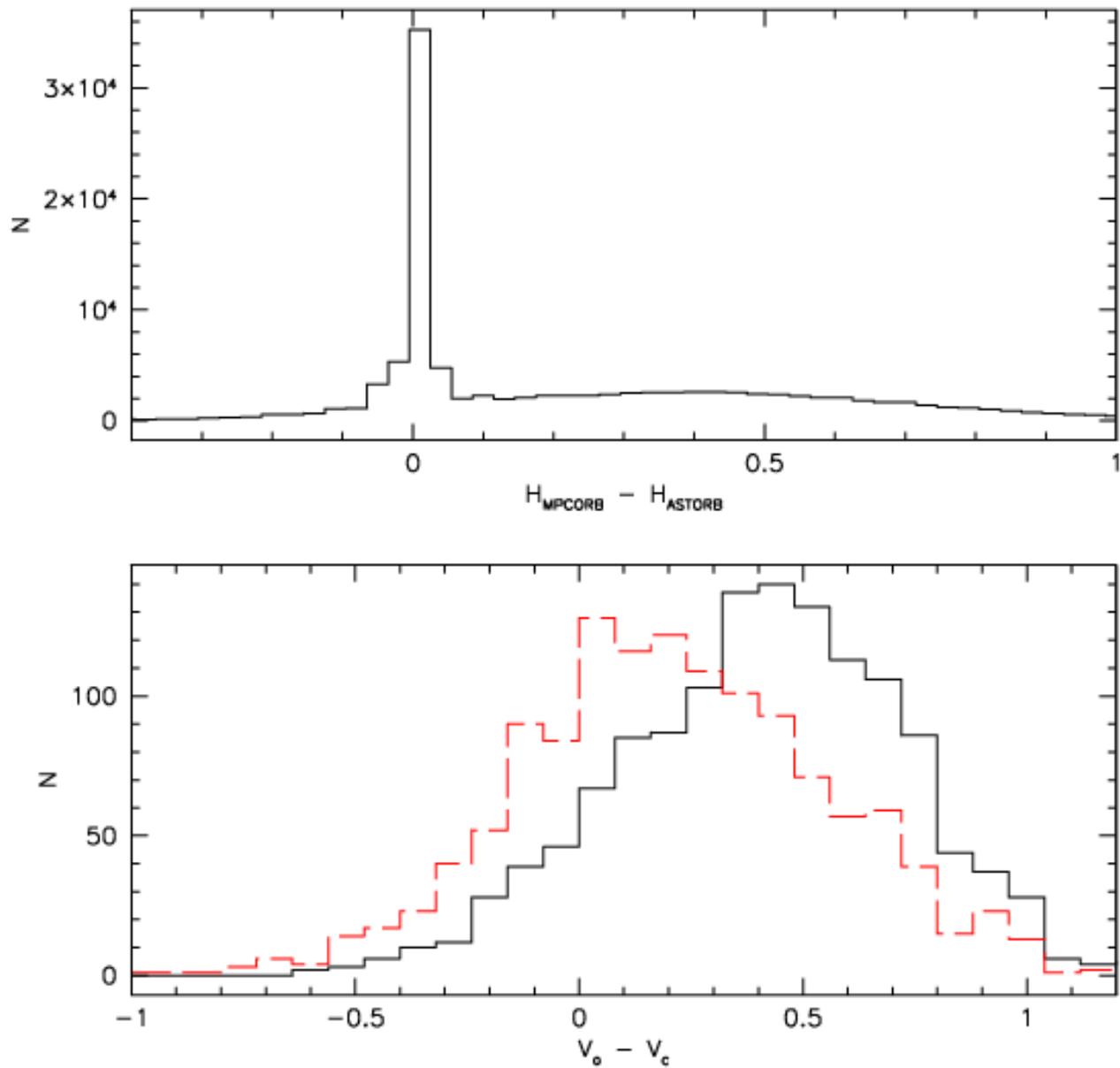}
\caption{
The top panel compares the difference in absolute magnitudes for the
115,583 common objects listed in the MPC and ASTORB catalogs. The bottom
panel compares the difference between the magnitude observed by SDSS
and predicted magnitudes for the ASTORB catalog (full line), and for
the MPC catalog (dashed line).
\label{m.mdiff_cats}
}
\end{figure}

It is of obvious interest to find out whether the offset between predicted
and observed magnitudes is particular to the ASTORB catalog, or perhaps also
present in other available databases. For this test we chose the Minor Planet
Center Orbit Database from September 5, 2001. It contains similar data to the
ASTORB catalog for 115,797 objects; 115,583 of these objects are common to
both catalogs. The top panel in Figure \ref{m.mdiff_cats} compares the
listed absolute magnitudes for the objects in common. As evident, the
catalogs contain two ``populations'' of object. For about half of the sample
the difference between the two values is less than 0.1 mag, while for the
remaining 57,719 objects it is larger than 0.1. The median value for the latter
is \about 0.4. We did not find any correlation between the difference in
absolute magnitude and other cataloged quantities. In particular, there
is no correlation between this difference and the mean absolute magnitude.
Given only two catalogs, it is not possible to tell which one is
responsible for the offset in magnitude scale. Fortunately, the catalogs can
be compared to SDSS observations of the matched objects for which
the photometric errors are not larger than 0.05 mag. The bottom panel
in Figure \ref{m.mdiff_cats} compares the $V_o - V_c$ histograms obtained
with the data from the ASTORB catalog (full line) and from the MPC catalog
(dashed line). As evident, the median discrepancy between the cataloged magnitudes
and the magnitudes measured by SDSS is about twice as large for the ASTORB
catalog as for the MPC catalog (\about0.4 mag vs. \about0.2 mag). Based on
this finding alone, it would seem that the MPC catalog should be used for
the matching purposes. However, the MPC catalog contains fewer objects than
the ASTORB catalog; out of the 1633 matched objects only 1387 are found in
the MPC catalog (implying an upper limit on the completeness of the MPC catalog
of 85\%).

The discrepancy in the absolute magnitude scale significantly affects
the number of objects brighter than a given limit. Assuming that the number
counts follow a $\log(N) = C + 0.6 \, V$ relation (I01), the MPC catalog will
imply 1.32 more objects than SDSS measurements, and the ASTORB catalog will
imply 1.74 more objects than SDSS measurements.

\begin{figure}
\plotone{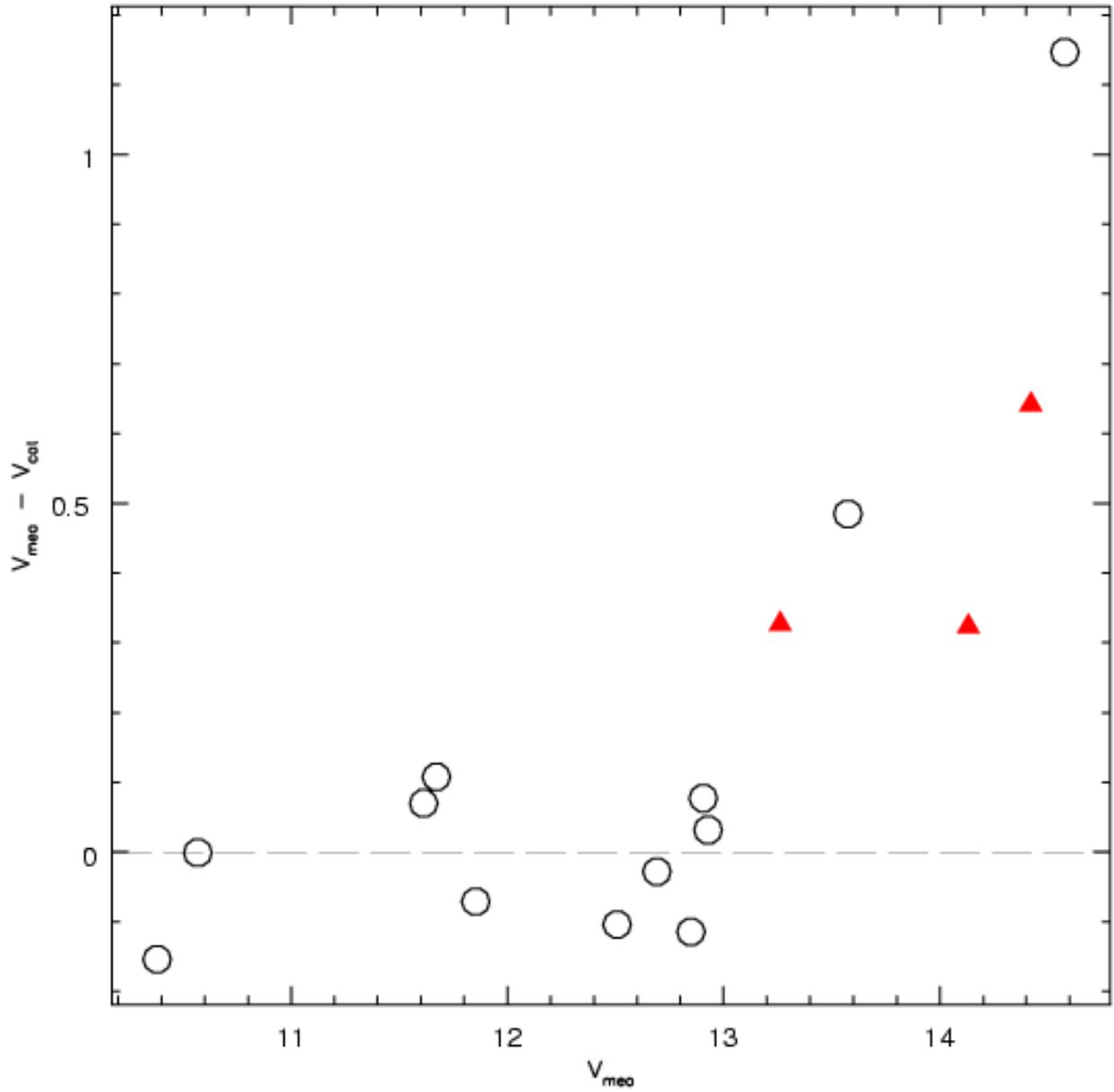}
\caption{
Differences between observed and calculated apparent V-band magnitudes of 15
asteroids observed by Krisciunas {\em et al.}. Asteroids having absolute
magnitudes accurately determined (given to two decimal places in ASTORB
catalog) are marked by circles. Triangles represent asteroids with poorly
known absolute magnitudes (given to one decimal place). Dashed horizontal
line serves as a visual aid, marking the position of zero magnitude
difference.
\label{fig_krisc}
}
\end{figure}

Although the uncertainty in the values of $V_o$ synthesized from the SDSS
photometry is probably smaller than 0.05 mag, it is prudent to verify the
offset in magnitude scales using non-SDSS observations. Krisciunas, Margon
and Szkody (1998) obtained photometric measurements in SDSS-like bands for 15
bright asteroids. Table 3 lists their measurements, and the synthesized,
$V_o$, and predicted, $V_c$, magnitudes. The $V_o - V_c$ difference is shown
as a function of $V_o$ in Figure \ref{fig_krisc}. Asteroids with $H$ in the
ASTORB catalog given to two decimal places are marked as circles, and those
with one decimal place are marked by triangles. The former are predominantly
at the bright end where they scatter around zero offset, indicating that the
photometric transformations from SDSS to Johnson V band are not biased.
The latter are found at the faint end and indicate a similar offset in the
magnitude scales as discussed above.

The correlation between the magnitude offset and the asteroid color
indicates that the discrepancy is probably due to differences in the
standard Johnson V magnitude and the ``visual'' magnitude listed in the
ASTORB catalog. A most likely cause is the fact that the majority of current
asteroid surveys (which produced most of the entries in the catalogs) are
obtained with open, unfiltered CCDs since they are mostly concentrated on
astrometric measurements. Thus, the measured
magnitudes are not equivalent to the standard Johnson V band, and
furthermore, the discrepancy depends on the asteroid color. Indeed, as 
discussed by Jedicke, Larsen \& Spahr (2002), the reporting of observed 
asteroid magnitudes is notoriously inconsistent. For example, the Minor 
Planet Center regularly weights magnitudes reported by different groups 
according to their historical photometric accuracy and transforms them 
to the $V$ band, while the ASTORB database uses the unweighted photometry 
in observed bands in determining the absolute magnitude.

\begin{figure}
\plotone{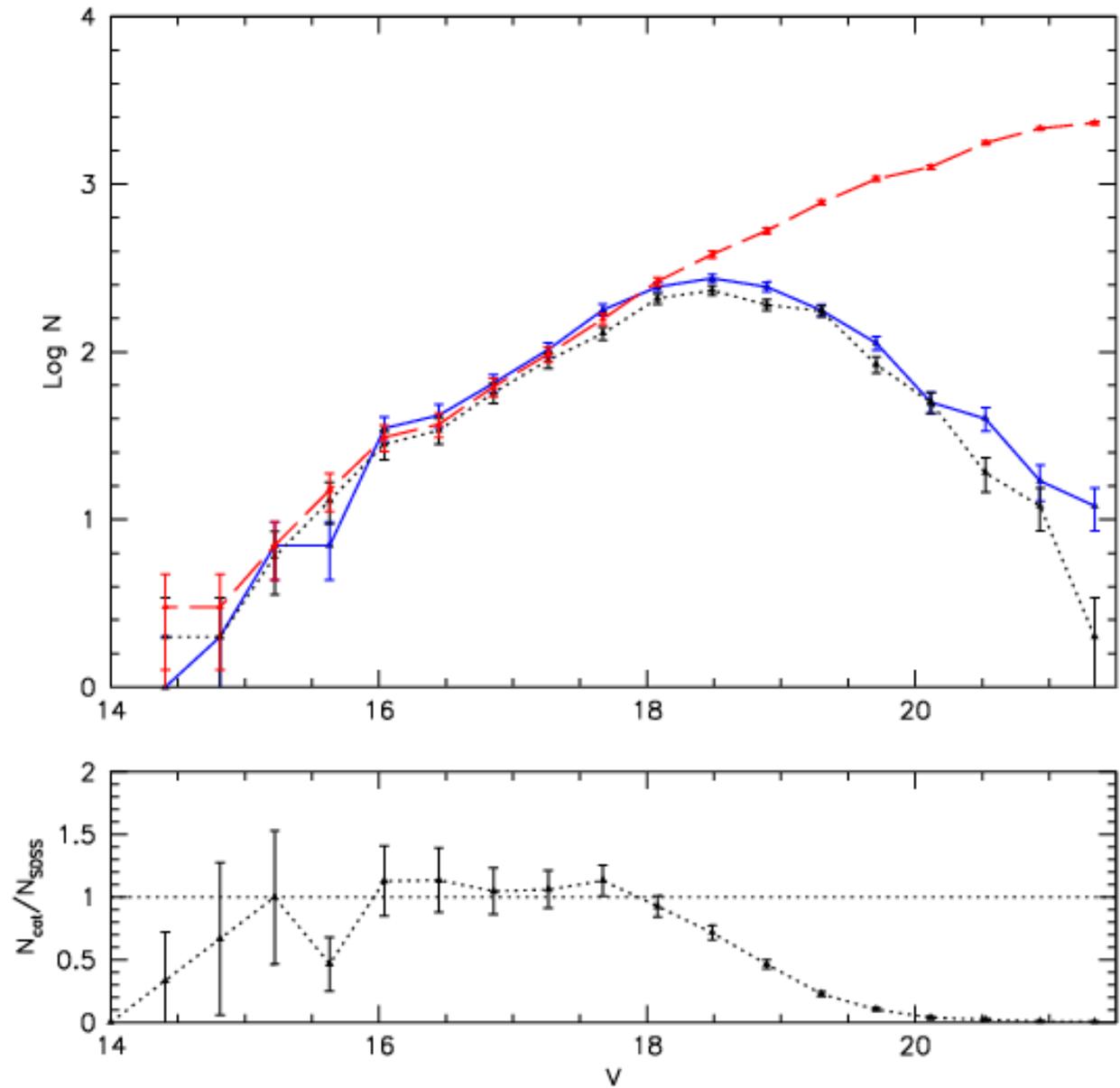}
\caption{The apparent magnitude distributions as in Figure \ref{fig_vlog},
except that the predicted magnitudes for ASTORB asteroids were offset by
0.41 mag (the full line). The shifted distribution implies SDSS completeness
in agreement with that shown in Figure \ref{fig_hlog_det}.
\label{fig_vlog_mod}
}
\end{figure}

Correcting the predicted magnitudes for a median offset of 0.41 mag shifts 
the $V_c$ histogram for ASTORB asteroids shown in Figure \ref{fig_vlog}
by the right amount to make the matched fraction consistent with the
estimate shown in Figure \ref{fig_hlog_det}. We show the shifted histogram
in Figure \ref{fig_vlog_mod}.

\section{            The Color Segregation in the Asteroid Belt              }
\label{colseg}

As discussed in Section \ref{sdssdata}, we joined the 1335 matched objects
from the SDSS Early Data Release with 1306 matched objects from unreleased runs, 
to form the final sample of 2641 unique matched objects. This is the sample used 
in our analysis of the color segregation in the asteroid belt.

\subsection{             Asteroid dynamical families                         }

Asteroid dynamical families are groupings of asteroids in proper\footnote{Proper
orbital elements are nearly invariants of motion and thus are well suited for
discovering objects with common dynamical history. They are different from the
osculating orbital elements which describe the current motion. For more details
see Milani \& Kne\v{z}evi\'{c} 1992, and references therein.} orbital
elements space, widely thought to be the results of collisional disruptions of
parent bodies (larger asteroids). Pioneered by Hirayama (1918),
the number and understanding of asteroid families has increased dramatically thanks
to a number of factors -- a large increase in the number of discovered asteroids,
more advanced knowledge of celestial mechanics and more sophisticated cluster-discovery
techniques. For a detailed review see Chapman {\em et al.} 1989. 

In a recent study, Zappala {\em et al.} (1995) identified 32 families and
31 ``clumps" using a hierarchical clustering method in a sample of 12,487 asteroids.
However, even a rough taxonomic classification exists only for a small number of these
objects. The knowledge of the chemical composition of each family could provide
further clues about the physical properties of the parent body. SDSS photometric
data can remedy this problem because the asteroids segregate in  SDSS multi-dimensional
color space. In this section we analyze the correlation between SDSS colors and the
distribution in orbital parameter space for 2641 matched asteroids. 

In order to study the correlation between SDSS colors and the various taxonomic
systems, we have cross-referenced the 1335 matched asteroids from the SDSS EDR sample
with a catalog of asteroids with known taxonomy obtained at the Planetary Data System
Small Bodies Node. The catalog contains information on 1199 objects, with
Tholen (978 objects), Barucci (438 objects), Tedesco (357 objects), Howell (112 objects)
and Xu (221 objects) taxonomies. There are only seven asteroids with taxonomical
information present in the SDSS sample, and are listed in Table 4. This small number 
is not surprising given that the SDSS saturation limit ($V \about 14$) is usually 
below the faint limit for most taxonomy surveys. The seven asteroids have SDSS colors 
in agreement with their taxonomic types (I01).

\subsection{           The Color properties of Asteroid Families             }

The 2641 known asteroids observed by SDSS have been cross-referenced with the
catalog of proper orbital elements produced by Milani {\em et al.} (1999) and
distributed through AstDyS. The resulting sample includes 1768 asteroids with
available proper elements. Hereafter we separate this sample into 531 blue
asteroids ($a^* < 0$, see I01), and 1237 red asteroids ($a^* > 0$). As discussed
by I01, there is a possibility that red asteroids with $i^*-z^* < -0.25$ form
a separate group in color-color space, and we treat separately the 131 asteroids
with such colors. In the remainder of this Section, we analyze the differences
in the distributions of these subsamples in the three-dimensional space spanned
by the semi-major axis, $a'$, inclination, $i'$, and eccentricity, $e'$.

\begin{figure}
\plotfiddle{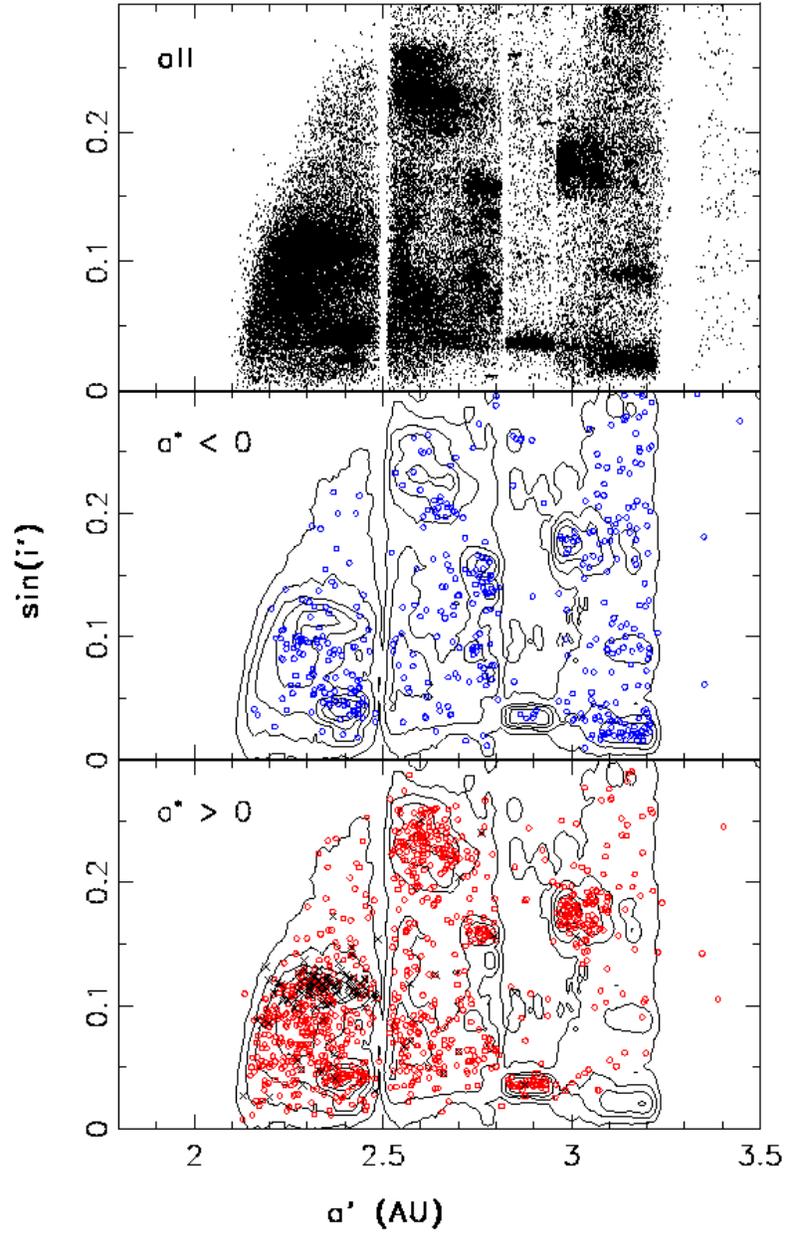}{14cm}{0}{65}{65}{-200}{-20}
\caption{The distribution of asteroids in the $\sin(i')$ vs. $a'$ plane. The top
shows 67,917 asteroids with the known proper orbital elements, marked as dots.
The remaining two panels show the same distribution as isodensity contours, with
the matched asteroids shows by open circles. The middle panel shows blue asteroids
(\a $<$ 0) and the bottom panel shows red asteroids (\a $>$ 0). A subset of red
asteroids with \iz $< -0.25$ is shown by crosses; most are found around $a'$ \about
2.2--2.5 and $\sin(i')$ \about 0.12.
\label{fig_a_i}
}
\end{figure}

\begin{figure}
\plotfiddle{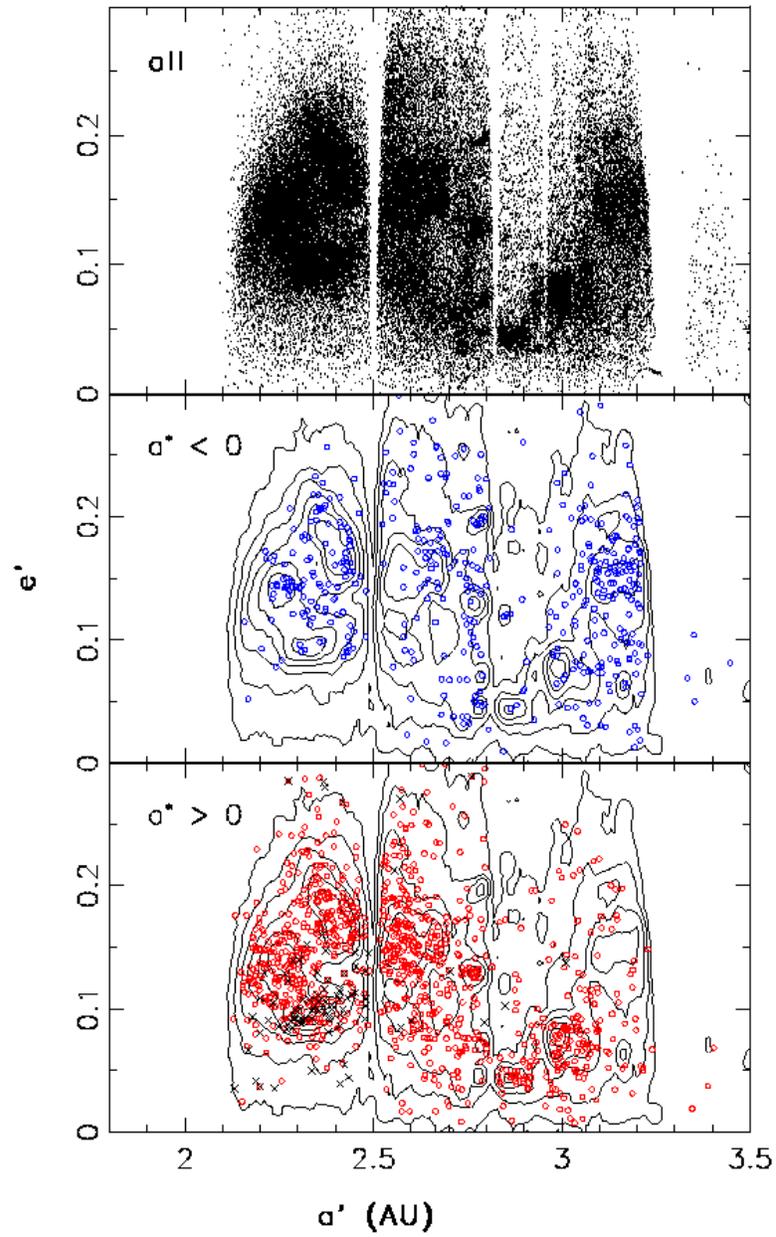}{14cm}{0}{65}{65}{-200}{-20}
\caption{Same as the previous figure, except that the distribution
of asteroids is shown in the $e'$ (eccentricity) vs. $a'$ plane.
\label{fig_a_e}
}
\end{figure}

\begin{figure}
\plotfiddle{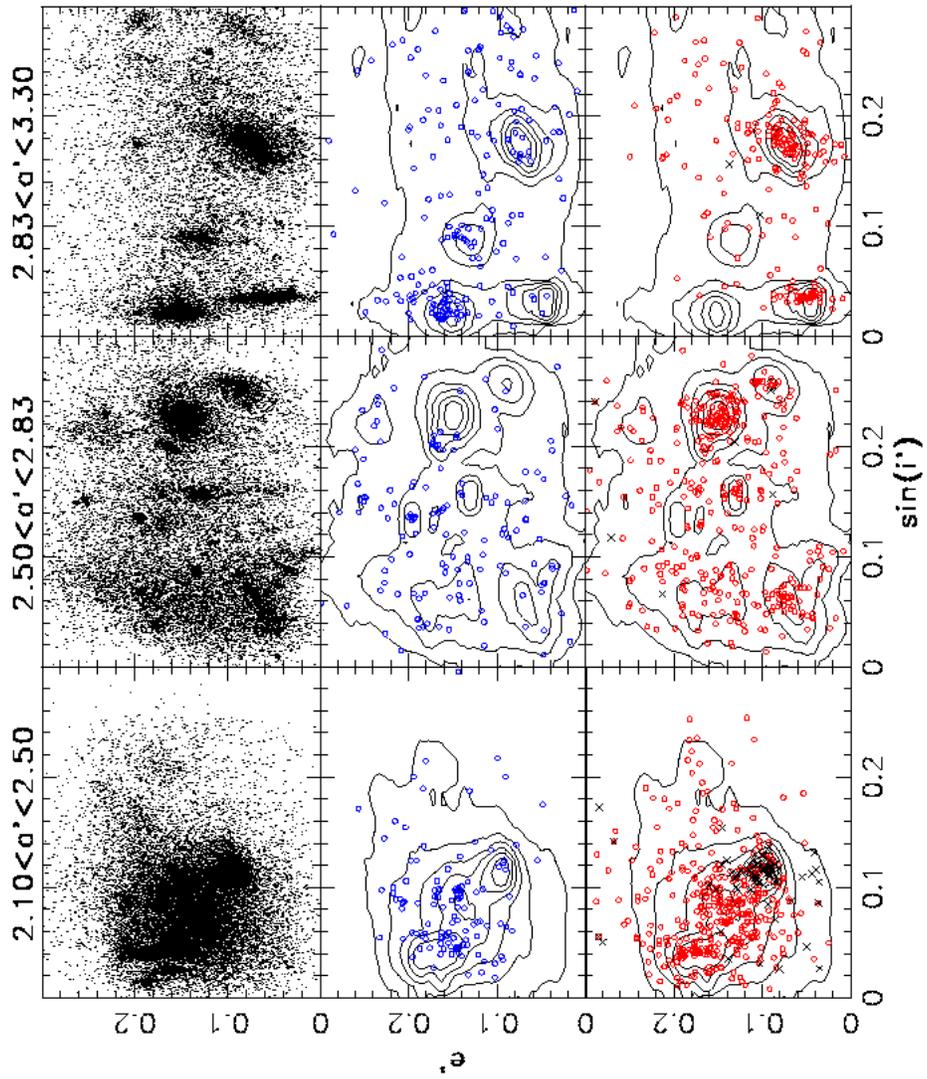}{14cm}{0}{65}{65}{-200}{-20}
\caption{Same as the previous figure, except that the distribution
of asteroids is shown in the $e'$ vs. $\sin(i')$ plane, and for three
ranges of the semi-major axis, $a'$, as marked on top.
\label{fig_i_e}
}
\end{figure}

Figure \ref{fig_a_i} displays the distribution of asteroids in the $a'$ vs.
$\sin(i')$ plane. The top panel shows 67,917 asteroids from the ASTORB database
which have known proper orbital elements (from Milani {\em et al.} 1999), marked as 
  dots. The major dynamical families (also known as the
Hirayama families) Eos, Koronis and Themis, are clearly visible at $a' > 2.8$ AU;
their approximate ($a', \sin(i')$) positions are (3.0, 0.18), (2.9, 0.03) and 
(3.15,0.02), respectively.
The 4:1, 3:1, 5:2 and 2:1 mean motion resonances with Jupiter at 2.065, 2.501, 
2.825 and 3.278 AU (the latter three
correspond to the Kirkwood gaps) are also evident. The $\nu_6$ resonance is
visible as a strong cuttof of asteroid density at high inclinations in
the $2.065 < a' < 2.501$ region. Due to using a sample which is about five times 
larger, the various families are more easily discerned than in Zappala {\em et al.} 
(1995); however, the distribution shown here agrees with their results.
The remaining two panels show the same distribution as isodensity contours, with
the matched asteroids shows by open circles. The middle panel shows blue asteroids
(\a $<$ 0), and the bottom panel shows red asteroids (\a $>$ 0). A subset of red
asteroids with \iz $< -0.25$ is shown by crosses; most are found around $a'$ \about
2.2--2.5 and $\sin(i')$ \about 0.12.
Figures \ref{fig_a_e} and \ref{fig_i_e} are analogous to Figure \ref{fig_a_i},
and show the asteroid distributions in the $e'$ vs. $a'$ and $e'$ vs. $\sin(i')$
planes. Figure \ref{fig_i_e} shows the distributions in three belt regions selected
by the positions of resonances.

For studying the chemical composition of various families, we adopt the family
definitions by Zappala {\em et al.} (1995). The distribution of blue and red asteroids
in the $a'$-$\sin(i')$-$e'$ space confirms the earlier results for the three largest
families. Koronis and Eos family seem to be mostly composed of red asteroids: 40 out
of 45 (89\%) asteroids associated with the Koronis family are red, and 77 of 98 (79\%)
are red for the Eos family. On the other hand, the Themis family is predominantly blue: 
41 out of 43 (95\%) members have $a^{*}<0$. All three major families are located in the 
outer belt. From other families in the outer belt we find two members of the Brasilia 
family, both are blue.

The middle portion of the belt, between the 3:1 and 5:2 resonances (2.5 $<a'<$ 2.8)
contains a number of smaller families. The SDSS sample discussed here provides good
color information for five of them. The Maria family is found to be red with 16 out of 
21 asteroids with $a^{*}>0$. The more numerous Eunomia family also displays red 
characteristics (97 red and 7 blue members).  On the other hand, the Adeona family and 
the Dora family are blue (12 out of 15, and all 11 members are blue, respectively). 
The Ceres family seems to be a mix of both asteroid types, with 11 blue and 24 red 
asteroids. 
Other smaller families have fewer identified members; we find marginal evidence that
the Hoffmeister, Rafita and Hestia families are predominantly red, while
the Misa and Taiyuan families are predominantly blue.

The inner part of the belt is dominated by the Vesta/Flora complex and at lower
inclinations by the Nysa/Polana and Massalia families. The Nysa/Polana group is found
to include a mix of red and blue asteroids, with red asteroids dominating the sample
in approximately 3:1 ratio (51 red and 19 blue). This is the strongest mixing of
blue and red asteroids found for the families studied here, and supports the claim by
Cellino {\em et al.} (2001) that the Nysa/Polana group includes two independent families, 
one composed of S (red) asteroids and the other including low-albedo F-like (presumably blue)
objects.

We note that for many families we find a substantial fraction ($\ga$ 10\%)
of the minor component (e.g. \about 20\% of the red Eos family members 
are blue). It is not easy to determine whether this mixing is due to the 
background contamination (i.e. by asteroids that are not family members), or 
due to non-homogeneous structure of the parent bodies\footnote{It is also 
possible that there is an asteroid color type that occupies the region 
around the boundary between blue and red types; e.g. see Figure 10 in I01}. 
A robust analysis 
necessarily involves precise definitions of families, and a careful study of the 
multi-dimensional color distribution for each family. Such an analysis
is beyond the scope of this work and will be presented in a future 
publication.

\subsection{            The V and J type Asteroids                    }

I01 analyzed the distribution of 316 asteroids in SDSS color space by
producing synthetic SDSS colors from the spectral measurements obtained by
the SMASS project (Xu {\em et al.} 1995). In addition to the two major
color types, there was an indication that red asteroids with extremely
blue \iz\ colors may form a separate class (see Figure 10 in I01). This
notion was also supported by the taxonomic classification of asteroids
with such colors; they were all classified as the J and V asteroids
associated with the Vesta family (Binzel \& Xu 1995). The proper orbital 
elements for 131 objects with \iz $< -0.25$ discussed here can be used
to test whether these objects cluster in the $a'$-$\sin(i')$-$e'$
space. 

\begin{figure}
\plotfiddle{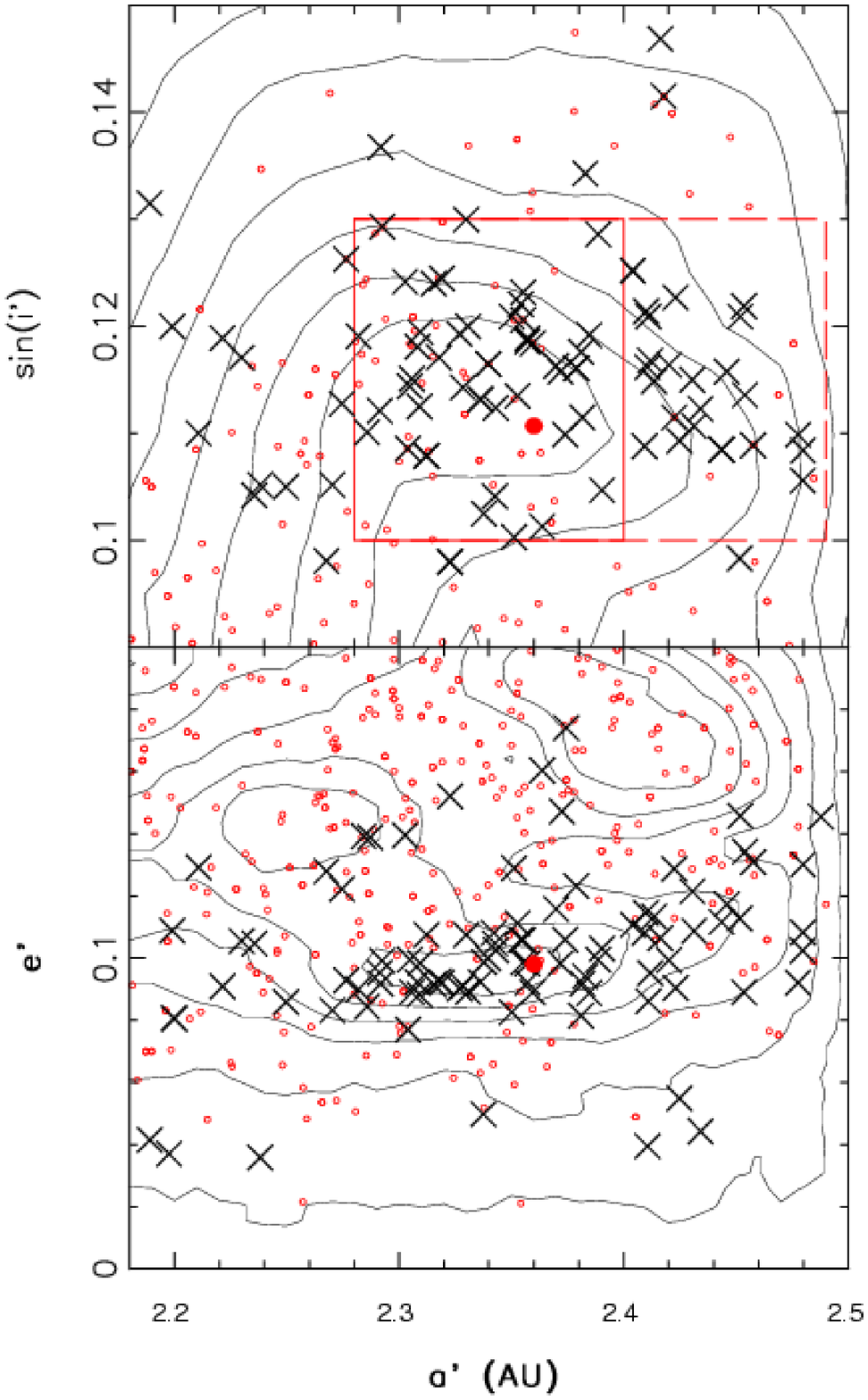}{14cm}{0}{65}{65}{-200}{-20}
\caption{The regions from the previous three figures that contain the Vesta
family. Only objects with \a $>$ 0 are shown; those with \iz $< -0.25$ are 
marked by crosses and others by circles. The boxes outline the core and
the tail regions, as discussed by Zappala {\em et al.} (1995). The position
of Vesta is marked by large dot.
\label{m.vestoids}
}
\end{figure}

The Vesta family occupies a very small volume in the orbital parameter space 
centered at $a'\about2.35$, $\sin(i')\about0.12$, and $e'\about0.1$. If the 
asteroids with extreme \iz\ colors can be interpreted as J and V taxonomic types,
then they should be concentrated in that region, rather than scattered
throughout the belt as blue and red asteroids are. The 131 red asteroids
with \iz $< -0.25$ are marked in Figures \ref{fig_a_i}, \ref{fig_a_e} and
\ref{fig_i_e} as crosses; it is evident that they are clustered around the
Vesta family. This is better seen in Figure \ref{m.vestoids} where we 
show only the small region of the orbital parameter space that is relevant 
for the Vesta family (2.1 $< a' <$ 2.5 and 0.09 $< \sin(i') <$0.15). The red
asteroids are shown as circles and the subset of 99 objects with \iz $< -0.25$ 
as crosses. The large dot marks the position of Vesta. The box outlined by 
the solid line is the core region (2.28 $< a' <$ 2.4 and 0.1 $< \sin(i') <$0.13), 
and the box outlined by the dashed line is the tail region (2.4 $< a' <$ 2.49) 
and 0.1 $< \sin(i') <$0.13), as proposed by Zappala {\em et al.} (1995).
Since a very high fraction of objects with \iz $< -0.25$ (99 out of 131 in the
displayed region, and 50 and 29, respectively, in the outlined core and tail 
regions) is found in the region of the orbital parameter space that is associated with 
the Vesta family, we conclude that the \iz\ color provides a reasonably reliable
method for selecting objects from the Vesta family.

\section{                           Discussion                  }

This work demonstrates the feasibility of significantly increasing the
number of asteroids with both accurate multi-color photometry and
known orbital elements by matching SDSS-detected asteroids with the 
catalogs of known asteroids. The sample discussed here includes \about 2600
asteroids; when SDSS is completed the final catalog could be more than
ten times larger. Such a large sample can be used for a compositional
analysis of the asteroid families at a level of detail which was not 
previously possible. We show here that SDSS color information is sufficient to
reveal strong color segregation among the asteroid families: most
are dominated by either blue or red objects.

We showed that the SDSS photometric pipeline automatically flags \about90\%
of observed moving objects, with a contamination level of only a few percent.
This is a better performance than for any other similar code.
The directly determined completeness is lower than the 98\% claimed by I01,
though without any significant consequences for their results.

It is somewhat surprising that only 66\% of the known asteroids listed in the ASTORB
catalog have sufficiently accurate orbital elements to predict their positions
to better than 30 arcsec. However, most objects with unreliable orbital
elements can be efficiently removed. It is much more perplexing that there is
a large offset in magnitude scales, not only with respect to SDSS, but also
among the standard asteroid databases. This offset results in an overestimate
of the number of asteroids brighter than a given magnitude limit by factor 1.7,
and resolves the discrepancy between the SDSS asteroid count normalization
and the number of asteroids listed in ASTORB catalog.

\vskip 0.4in
\leftline{Acknowledgments}

We thank Robert Jedicke for fruitful discussions regarding the discrepancy
in asteroid counts between SDSS and ASTORB database, and Dejan Vinkovi\'{c}
for helping with the visual inspection of SDSS images. We thank Princeton 
University for generous financial support of this research. 

The Sloan Digital Sky Survey (SDSS) is a joint project of The University of Chicago,
Fermilab, the Institute for Advanced Study, the Japan Participation Group, The Johns
Hopkins University, the Max-Planck-Institute for Astronomy, the Max-Planck-Institute
for Astrophysics, New Mexico State University, Princeton University, the United States
Naval Observatory, and the University of Washington. Apache Point Observatory, site
of the SDSS telescopes, is operated by the Astrophysical Research Consortium (ARC).

Funding for the project has been provided by the Alfred P. Sloan Foundation,
the SDSS member institutions, the National Aeronautics and Space Administration,
the National Science Foundation, the U.S. Department of Energy, Monbusho, and the
Max Planck Society. The SDSS Web site is http://www.sdss.org/.


\newpage
\clearpage

\begin{scriptsize}
\begin{deluxetable}{crrrrr}
\tablenum{1}
\label{tblMatching}
\tablecolumns{6}
\tablewidth{340pt}
\tablecaption{All ASTORB asteroids matched to SDSS objects}
\tablehead
{
    Run &  ASTORB   &  SDSS   & Matched &  Missed by  & Not found \\
        & asteroids & objects & objects & {\it photo} &   ASTORB
}
\startdata
  94    &     663   &   2757  &   395   &      94     &     174   \\
 125    &     648   &   2844  &   426   &      48     &     174   \\
 752    &     712   &   3422  &   373   &      33     &     306   \\
 756    &     778   &   3645  &   439   &      44     &     295   \\
\hline
  All   &    2801   &  12668  &  1633   &     219     &     949   \\
 Unique &    2170   &         &  1325   &             &           \\
\enddata
\end{deluxetable}
\end{scriptsize}

\begin{scriptsize}
\begin{deluxetable}{crrrrrr}
\tablenum{2}
\label{tblMatchingConstrained}
\tablecolumns{6}
\tablewidth{340pt}
\tablecaption{Reliable ASTORB Asteroids Matched to SDSS objects}
\tablehead
{
    Run &  ASTORB   &  SDSS   & Matched &  Missed by  & Not found \\
        & asteroids & objects & objects & {\it photo} &   ASTORB
}
\startdata
  94    &     391   &   2757  &  290    &    82       &    19     \\
 125    &     402   &   2844  &  350    &    32       &    20     \\
 752    &     374   &   3422  &  318    &    25       &    31     \\
 756    &     445   &   3645  &  377    &    34       &    34     \\
\hline
  All   &    1612   &  12668  & 1335    &   173       &   104     \\
 Unique &    1239   &         &  960    &             &           \\
\enddata
\end{deluxetable}
\end{scriptsize}

\begin{scriptsize}
\begin{deluxetable}{rlcccccccr}
\label{tbl_krisc}
\tablenum{3}
\tablewidth{0pc}
\tablecaption{The Photometric Observations of Asteroids$^a$}
\tablehead{
\colhead{No.} & \colhead{Name} & \colhead{Type} &
\colhead{1997 Date} & \colhead{UT} &
\colhead{r$^*$} &
\colhead{g$^*-$r$^*$} & \colhead{V$_{mea}^b$} &
\colhead{V$_{cat}^c$} & \colhead{V$_{mea}$-V$_{cat}$}
 }
\startdata
446 & Aeternitas & A  & 30 Sep & 11:09 & 14.33 & 0.61 & 14.58 & 13.43 &  1.14\phantom{xxx} \nl
702 & Alauda     & C  & 30 Sep & 11:47 & 12.75 & 0.45 & 12.93 & 12.90 &  0.03\phantom{xxx} \nl
82  & Alkmene    & S  & 28 Sep & 10:57 & 12.43 & 0.64 & 12.69 & 12.72 & -0.02\phantom{xxx} \nl
774 & Armor      & S  & 28 Sep & 03:52 & 12.97 & 0.71 & 13.26 & 12.94 &  0.32\phantom{xxx} \nl
371 & Bohemia    & AS &  4 Oct & 05:05 & 12.64 & 0.65 & 12.90 & 12.83 &  0.07\phantom{xxx} \nl
\nl
349 & Dembowska  & R  & 29 Sep & 11:54 & 10.28 & 0.70 & 10.57 & 10.57 & -0.00\phantom{xxx} \nl
433 & Eros$^b$   & S  & 30 Sep & 12:26 & 13.30 & 0.67 & 13.57 & 13.09 &  0.48\phantom{xxx} \nl
480 & Hansa      & S  & 29 Sep & 09:59 & 12.25 & 0.63 & 12.51 & 12.61 & -0.10\phantom{xxx} \nl
10  & Hygeia     & C  & 28 Sep & 12:05 & 11.43 & 0.46 & 11.61 & 11.54 &  0.06\phantom{xxx} \nl
683 & Lanzia     & C  &  1 Oct & 03:35 & 14.23 & 0.48 & 14.42 & 13.78 &  0.64\phantom{xxx} \nl
\nl
68  & Leto       & S  & 27 Sep & 10:37 & 10.11 & 0.66 & 10.38 & 10.53 & -0.15\phantom{xxx} \nl
149 & Medusa     & S  & 29 Sep & 06:12 & 12.57 & 0.68 & 12.85 & 12.96 & -0.11\phantom{xxx} \nl
196 & Philomela  & S  & 28 Sep & 08:51 & 11.41 & 0.64 & 11.67 & 11.56 &  0.10\phantom{xxx} \nl
314 & Rosalia    & C  &  5 Oct & 03:15 & 13.95 & 0.46 & 14.13 & 13.81 &  0.32\phantom{xxx} \nl
138 & Tolosa     & S  &  5 Oct & 04:11 & 11.57 & 0.69 & 11.85 & 11.92 & -0.07\phantom{xxx} \nl
\nl
\enddata
\tablenotetext{a}{Observations with SDSS-like filters by Krisciunas, Margon
                  \& Szkody (1998).}
\tablenotetext{b}{The synthesized Johnson V magnitude based on observations
                  using SDSS-like filters.}
\tablenotetext{c}{The predicted catalog-based Johnson V magnitude.}
\end{deluxetable}
\end{scriptsize}

\begin{scriptsize}
\begin{deluxetable}{rrrr}
\tablenum{4}
\label{tbl_taxonomy}
\tablecolumns{4}
\tablewidth{240pt}
\tablecaption{Asteroids with known taxonomy in the SDSS EDR sample}
\tablehead
{
       Asteroid     &  Taxonomy$^a$   & $a^{*}$ & $i^*-z^*$\\
}
\startdata
( 220) Stephania    & XC (Tholen) & -0.062  &  -0.012  \\
(1628) Strobel      & X (SMASS)   &  0.024  &   0.079  \\
(1679) Nevanlinna   & X (SMASS)   & -0.048  &   0.036  \\
(1711) Sandrine     & S (Tholen)  &  0.074  &   0.042  \\
(1772) Gagarin      & S (SMASS)   &  0.135  &   0.003  \\
(2215) Sichuan      & S (SMASS)   &  0.097  &   0.011  \\
(5118) Elnapoul     & S (SMASS)   &  0.151  &  -0.022  \\
\enddata
\tablenotetext{a}{The asteroid taxonomic classification and its source.}
\end{deluxetable}
\end{scriptsize}

\newpage

\clearpage

\clearpage

\begin{thebibliography}{99}

\bibitem[Binzel \& Xu 1993]{BinzelXu93} Binzel, R.P., Xu, S. 1995, Science, 260, 186

\bibitem[Bowell 1989]{Bowell89}
    Bowell, E., {\em et al.} 1989, in Asteroids II, eds. R.P. Binzel,
        T. Gehrels and M.S. Matthews, (Tucson: Univ. of Arizona Press), 549

\bibitem[Bowell 2001]{AstorbURL} Bowell, E. 2001, Introduction to ASTORB,
    available from ftp://ftp.lowell.edu/pub/elgb/astorb.html

\bibitem[Cellino {\em et al.} 2001]{C01} Cellino, A., {\em et al.}, Icarus, 152, 225

\bibitem[Chapman 1989]{Chapman89} Chapman, C.R., {\em et al.} 1989, in Asteroids II, 
    eds. R.P. Binzel, T. Gehrels and M.S. Matthews, (Tucson: Univ. of Arizona Press), 549

\bibitem[Fukugita {\em et al.} 1996]{F96}Fukugita, M., Ichikawa, T., Gunn, J.E.,
         Doi, M., Shimasaku, K., \& Schneider, D.P. 1996, AJ, 111, 1748

\bibitem[Gradie, Chapman \& Williams 1988]{GCTW79} Gradie, J., Chapman, C.R., \& Williams, 
        J.G. 1979, in Asteroids, ed. T. Gehrels, (Tuscon: Univ. of Arizona Press), 359

\bibitem[Ivezi\'c {\em et al.} 2001]{Ivezic01} Ivezi\'c, \v Z., {\em et al.}
         2001, Astronomical Journal, 122, 2749.

\bibitem[Jedicke, Larsen \& Spahr 2002]{JLS02} Jedicke, R., Larsen, J.
         \& Spahr, T. 2002, {\it Observational Selection Effects in 
         Asteroid Surveys and Estimates of Asteroid Population Sizes}, 
         Proceedings of the Asteroids 2001 ``From Piazzi to the 3$^{\rm rd}$ 
         Millennium'' Conference, June 11--16, 2001, Palermo, Italy, 
         University of Arizona Press, in press.

\bibitem[Juri\'c \& Korlevi\'c 2000]{Juric00} Juri\'c M., and Korlevi\'c K. 2000,
        Vi\v{s}njan Observatory Image Database 2, Proceedings of the CARNet Users
        Conference 2000, Zagreb

\bibitem[Krisciunas, Margon \& Szkody 1998]{KMS98} Krisciunas, K., Margon,
        B., \&  Szkody P. 1998, PASP 110, 1342

\bibitem[Lupton {\em et al.} 2002]{Lupton02} Lupton, R.H., {\em et al.} 2002,
         in preparation

\bibitem[Milani {\em et al.} 1999]{Milani99} Milani, A., {\em et al.} 1999,
         Icarus, 137, 269

\bibitem[Milani \& Kne\v{z}evi\'{c} 1992]{Milani92} Milani, A. \& Kne\v{z}evi\'{c},
         Z. 1992, Icarus, 98, 211.

\bibitem[Muinonen \& Bowell 1993]{MB93} Muinonen, K., \& Bowell, E.
        1993, Icarus, 104, 255

\bibitem[DE405]{DE405} Standish, E. M. 1998, JPL Planetary and Lunar Ephemerids,
    available at http://ssd.jpl.nasa.gov/eph\_info.html

\bibitem[Stoughton {\em et al.} 2002]{SDSSEDR} Stoughton, C., {\em et al.}
         2002, Astronomical Journal, in press

\bibitem[Xu {\em et al.} 1995]{Xu95} Xu, S., Binzel, R.P., Burbine, T.H. \& Bus,
         S.J. 1995, Icarus, 115, 1

\bibitem[Zappala {\em et al.} 1995]{Zappala95} Zappala, V., Bendjoya, Ph.,
         Cellino, A., Farinella, P., \& Froeschle, C. 1995, Icarus, 116, 291

\end{thebibliography}
\end{document}